\documentclass[twocolumn,]{aastex63}

\hypersetup{linkcolor=brown, filecolor=lightgray, urlcolor=olive}

\usepackage{graphicx,amsmath,amssymb,amstext,natbib}
\usepackage{epsf}
\usepackage{epstopdf}
\usepackage{xspace}

\graphicspath{{./}{figures/}}
\accepted{January 11, 2021}

\shorttitle{The history of cosmic star formation}
\shortauthors{J. A. Zavala et al.}

\begin{document}
\title{\large The Evolution of the IR Luminosity Function and Dust-obscured Star Formation in the Last 13 Billion Years}

\correspondingauthor{Jorge A. Zavala}
\email{jzavala@utexas.edu}

\author[0000-0002-7051-1100]{J. A. Zavala}
\affil{The University of Texas at Austin, 2515 Speedway Blvd Stop C1400, Austin, TX 78712, USA}

\author[0000-0002-0930-6466]{C. M. Casey}
\affil{The University of Texas at Austin, 2515 Speedway Blvd Stop C1400, Austin, TX 78712, USA}

\author[0000-0003-0415-0121]{S. M. Manning}
\affil{The University of Texas at Austin, 2515 Speedway Blvd Stop C1400, Austin, TX 78712, USA}

\author[0000-0002-6290-3198]{M. Aravena}
\affil{N\'ucleo de Astronom\'ia, Facultad de Ingenier\'ia y Ciencias, Universidad Diego Portales, Av. Ej\'ercito 441, Santiago, Chile. }

\author[0000-0002-3915-2015]{M. Bethermin}
\affil{Aix Marseille Univ, CNRS, LAM, Laboratoire d'Astrophysique de Marseille, Marseille, France }

\author[0000-0001-8183-1460]{K. I. Caputi}
\affil{Kapteyn Astronomical Institute, University of Groningen, P.O. Box 800, 9700AV Groningen, The Netherlands }
\affil{Cosmic Dawn Center (DAWN)}

\author[0000-0002-9548-5033]{D. L. Clements}
\affil{Imperial College London, Blackett Laboratory, Prince Consort Road, London, SW7 2AZ, UK }

\author[0000-0001-9759-4797]{E. da Cunha}
\affil{International Centre for Radio Astronomy Research, University of Western Australia, 35 Stirling Hwy, Crawley, WA 6009, Australia }

\author[0000-0003-3627-7485]{P. Drew}
\affil{The University of Texas at Austin, 2515 Speedway Blvd Stop C1400, Austin, TX 78712, USA }

\author[0000-0001-8519-1130]{S. L. Finkelstein}
\affil{The University of Texas at Austin, 2515 Speedway Blvd Stop C1400, Austin, TX 78712, USA }

\author[0000-0001-7201-5066]{S. Fujimoto}
\affil{Cosmic Dawn Center (DAWN)}
\affil{Niels Bohr Institute, University of Copenhagen, Lyngbyvej 2, DK-2100 Copenhagen, Denmark }

\author[0000-0003-4073-3236]{C. Hayward}
\affil{Center for Computational Astrophysics, Flatiron Institute, 162 Fifth Avenue, New York, 10010, USA }

\author[0000-0001-6586-8845]{J. Hodge}
\affil{Leiden Observatory, Niels Bohrweg 2, 2333 CA Leiden, The Netherlands }

\author[0000-0001-9187-3605]{J. S. Kartaltepe}
\affil{School of Physics and Astronomy, Rochester Institute of Technology, 84 Lomb Memorial Drive, Rochester NY 14623, USA }

\author[0000-0002-7821-8873]{K. Knudsen}
\affil{Chalmers University of Technology, Department of Space, Earth and Environment, Onsala Space Observatory, 439 92 Onsala, Sweden }

\author[0000-0002-6610-2048]{A. M. Koekemoer}
\affil{Space Telescope Science Institute, 3700 San Martin Dr., Baltimore, MD 21218, USA }

\author[0000-0002-7530-8857]{A. S. Long}
\affil{Department of Physics and Astronomy, University of California, Irvine, CA 92697, USA }

\author[0000-0002-4872-2294]{G. E. Magdis}
\affil{Cosmic Dawn Center (DAWN)}
\affil{DTU-Space, Technical University of Denmark, Elektrovej 327, DK-2800 Kgs. Lyngby, Denmark Institute for Astronomy, Astrophysics, Space Applications and Remote Sensing, National Observatory of Athens, GR-15236 Athens, Greece }
\affil{Niels Bohr Institute, University of Copenhagen, Lyngbyvej 2, DK-2100 Copenhagen, Denmark }
\affil{Institute for Astronomy, Astrophysics, Space Applications and Remote Sensing, National Observatory of Athens, GR-15236 Athens, Greece}

\author[0000-0003-2475-124X]{A. W. S. Man}
\affil{Dunlap Institute for Astronomy and Astrophysics, University of Toronto, 50 St George Street Toronto ON, M5S 3H4, Canada }

\author[0000-0003-1151-4659]{G. Popping}
\affil{European Southern Observatory, Karl-Schwarzschild-Strasse 2, 85748, Garching, Germany }

\author[0000-0002-1233-9998]{D. Sanders}
\affil{Institute for Astronomy, 2680 Woodlawn Drive, University of Hawaii, Honolulu, HI 96822, USA }

\author[0000-0002-0438-3323]{N. Scoville}
\affil{California Institute of Technology, MC 249-17, 1200 East California Boulevard, Pasadena, CA 91125, USA }

\author[0000-0002-5496-4118]{K. Sheth}
\affil{NASA Headquarters, 300 E Street SW, Washington, DC 20546, USA }

\author[0000-0002-8437-0433]{J. Staguhn}
\affil{The Henry A. Rowland Department of Physics and Astronomy, Johns Hopkins University, 3400 North Charles Street, Baltimore, MD 21218, USA }
\affil{Observational Cosmology Lab, Code 665, NASA Goddard Space Flight Center, Greenbelt, MD 20771, USA }]

\author[0000-0003-3631-7176]{S. Toft}
\affil{Cosmic Dawn Center (DAWN)}
\affil{Niels Bohr Institute, University of Copenhagen, Lyngbyvej 2, DK-2100 Copenhagen, Denmark }

\author[0000-0001-7568-6412]{ E. Treister}
\affil{Instituto de Astrof\'isica, Facultad de F\'isica, Pontif\'icia Universidad Cat\'olica de Chile, Casilla 306, Santiago 22, Chile}

\author[0000-0001-7192-3871]{J. D. Vieira}
\affil{Department of Astronomy, University of Illinois, 1002 West Green St., Urbana, IL 61801 }

\author[0000-0001-7095-7543]{ M. S. Yun}
\affil{Department of Astronomy, University of Massachusetts, Amherst, MA 01003, USA }

\begin{abstract}
We present the first results from the 2\,mm Mapping Obscuration to Reionization (MORA) survey, the largest ALMA blank-field contiguous survey to-date (184\,arcmin$^2$) and the only at 2\,mm to search for dusty star-forming galaxies (DSFGs). We use the 13 sources detected above $5\sigma$ to estimate the first ALMA galaxy number counts at this wavelength. These number counts are then combined with the state-of-the-art galaxy number counts at 1.2\,mm and 3\,mm and with a backward evolution model to place constraints on the evolution of the IR luminosity function and dust-obscured star formation in the last 13 billion years. Our results suggest a steep redshift evolution on the space density of DSFGs and confirm the flattening of the IR luminosity function at faint luminosities, with a slope of $\alpha_{\rm LF}=-0.42^{+0.02}_{-0.04}$. We conclude that the dust-obscured component, which peaks at $z\approx2-2.5$, has  dominated  the  cosmic history  of  star  formation for the past $\sim12$ billion years, back to $z\sim4$.
At $z=5$, the dust-obscured star formation is estimated to be $\sim35\%$ of the total star formation rate density and decreases to $25\%-20\%$ at $z=6-7$, implying a minor contribution of 
dust-enshrouded star formation  in  the  first  billion years of the Universe. 
With the dust-obscured star formation history constrained up to the end of the epoch of reionization, our results provide a benchmark to test galaxy formation models, to study the galaxy mass assembly history, and to understand the dust and metal enrichment of the  Universe at early times.\\
\end{abstract}

\keywords{  Galaxy evolution --- High-redshift galaxies --- Star formation --- Galaxy counts --- Luminosity function --- Dust continuum emission\\} 

\section{Introduction} \label{secc:intro}
Mapping the cosmic history of star-formation is of fundamental importance in our understanding of galaxy formation and evolution  not only because it contains the galaxy mass assembly history but also because it represents the footprint of the metal enrichment of the Universe. 

A complete, unbiased determination of the star formation rate density (SFRD) requires a multi-wavelength approach
to directly probe the stellar emission of new-born stars
as well as the starlight that has been absorbed (then re-emitted) in dust-enshrouded regions. The former is efficiently measured using rest-frame ultraviolet (UV) observations and the latter using far-infrared (FIR) and (sub-)millimeter surveys that trace the dust re-processed  emission from young stars (see \citealt{Madau2014a} for a  review). 

The current UV census of star formation reaches out to $z\sim11$, close to the formation epoch of the first galaxies (\citealt{Oesch2018a}). Nevertheless, despite some individual detections of dust in galaxies up to $z\sim8$ (\citealt{Watson2015a,Laporte2017a}), studies of the global dust-obscured star formation rate density from FIR/sub-mm surveys are very limited at $z>3$.

The highest redshift  estimations of the SFRD from FIR-to-mm surveys, at $z=4.5-5.5$, are uncertain even at the $\approx1.0-1.5\,$dex level (\citealt{Rowan-Robinson2016a,Williams2019a,Loiacono2020a}) and differ by up to two order of magnitudes (\citealt{Michalowski2017a} cf. \citealt{Rowan-Robinson2016a}). This is mostly due to the low number statistics since most of the samples used to derive these measurements range from only one (\citealt{Williams2019a,Loiacono2020a}) to a handful of objects (\citealt{Dudzeviciute2020a,Gruppioni2020a}).
This implies that the total amount of  
dust-enshrouded star formation in the earliest epochs of the Universe has remained unknown and, therefore, our current picture of the history of cosmic star formation remains incomplete.

This study aims at estimating the 
history of dust-obscured star formation back to $z\sim7$ by combining a backward galaxy evolution model of the dusty star-forming galaxy (DSFG)
population (\citealt{Casey2018b,Casey2018a}) with the state-of-the-art Atacama Large Millimeter/submillimeter Array (ALMA)  surveys. 

We infer constraints on the prevalence and
characteristics of these galaxies through measurements of galaxy number counts at different
wavelengths. Despite its relative conceptual simplicity, number counts have proven to be a very powerful tool to test and constrain galaxy formation models (e.g. \citealt{Baugh2005a,Hayward2013a}), and consequently, represent an essential measurement in any long-wavelength survey. 

 
The galaxy number counts used in this work include the first arcsecond-resolution interferometric number counts at 2\,mm  achieved with ALMA and the most recent estimations at 1.2\,mm and 3\,mm. The former is from our new 2\,mm Mapping Obscuration to Reionization (MORA) survey described in \S\ref{sec:MORA}, the first ALMA large  map at this wavelength with a total area of 184\,arcmin$^2$ (an order of magnitude larger than previous interferometric blind surveys). At 3\,mm, the measurements come from the $\sim200\,$arcmin$^2$ ALMA  archival program reported in \cite{Zavala2018c}, which we revise in this work after finding three false detections in their sample (see \S\ref{secc:3mm_num_counts}). 
These number counts are also complemented with those from the 4.2\,arcmin$^2$ deep ALMA large program ASPECS survey (at 1.2\,mm and 3\,mm; \citealt{Gonzalez-Lopez2019a,Gonzalez-Lopez2020a}), in which 40 continuum sources were detected.  All  these surveys represent the state-of-the-art observations at long wavelengths, comprising  not only the deepest measurements
at those wavebands (beyond those assisted by gravitational amplification), but also the only interferometric surveys at 2 and 3\,mm, which are  crucial to constraining the infrared (IR) luminosity function (IRLF) at high-redshift (see discussion by \citealt{Bethermin2015a,Casey2018b,Casey2018a,Zavala2018c}, and \citealt{Wilner1997a} for a pioneering $\sim3\,$mm survey).



The constraints provided by these surveys and by all the far-infrared and sub-millimeter data aggregated over the  last two  decades, allow us to infer the  IR luminosity function and dust-obscured star formation in the last 13 billion years of the Universe, pushing the current redshift limits from $z\sim4-5$ to $z\sim7$. Our constraints are discussed in \S\ref{secc4} and summarized in \S\ref{secc:conclusions}.






Trough this work we assume $H_0=67.7\,\rm km\,s^{-1}\,Mpc^{-1}$ and $\Omega_{\lambda}=0.69$ (\citealt{Planck2016a}), and a Chabrier initial mass function (\citealt{Chabrier2003a}) for SFR estimations. 

\section{Mapping Obscuration to Reionization ALMA (MORA) Survey} \label{sec:MORA}

\begin{figure*}[t]\centering
\includegraphics[width=0.5\textwidth]{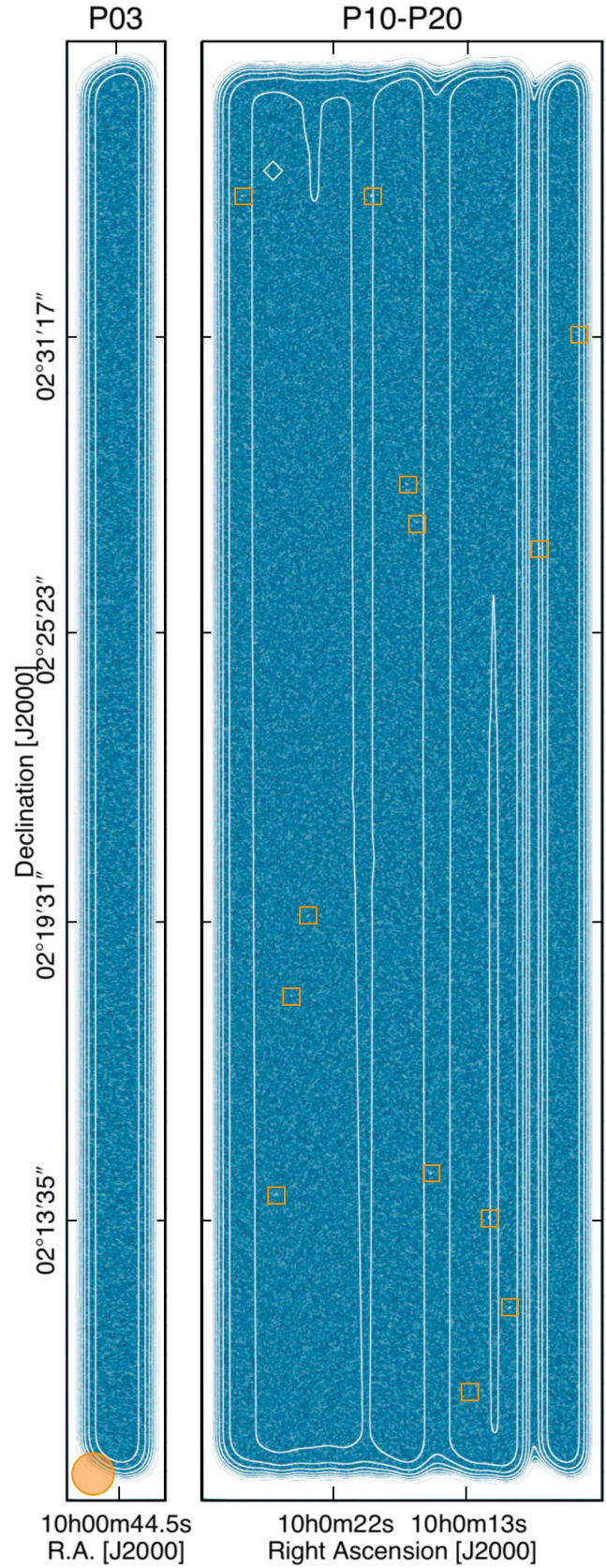}
\caption{The MORA 2\,mm SNR maps and the 13 sources detected above $5\sigma$ (orange squares). The noise variations across the mosaics are illustrated by the contours, which range from $60\rm\,\mu Jy$ to $240\rm\,\mu Jy$ in steps of $30\rm\,\mu Jy$ . The typical size of the primary beam response at the frequency  of our observations ($\theta_{\rm FWHM}\approx43''$) is represented by the  orange circle (note that the synthesized beamsize of $\theta_{\rm FWHM}\sim1.5\arcsec$ is significantly smaller).  The $-6\sigma$ peak is also identified with the pale green diamond. 
\label{fig:map}}
\end{figure*}

\begin{figure}[t]\hspace{-0.6cm}
\includegraphics[width=0.53\textwidth]{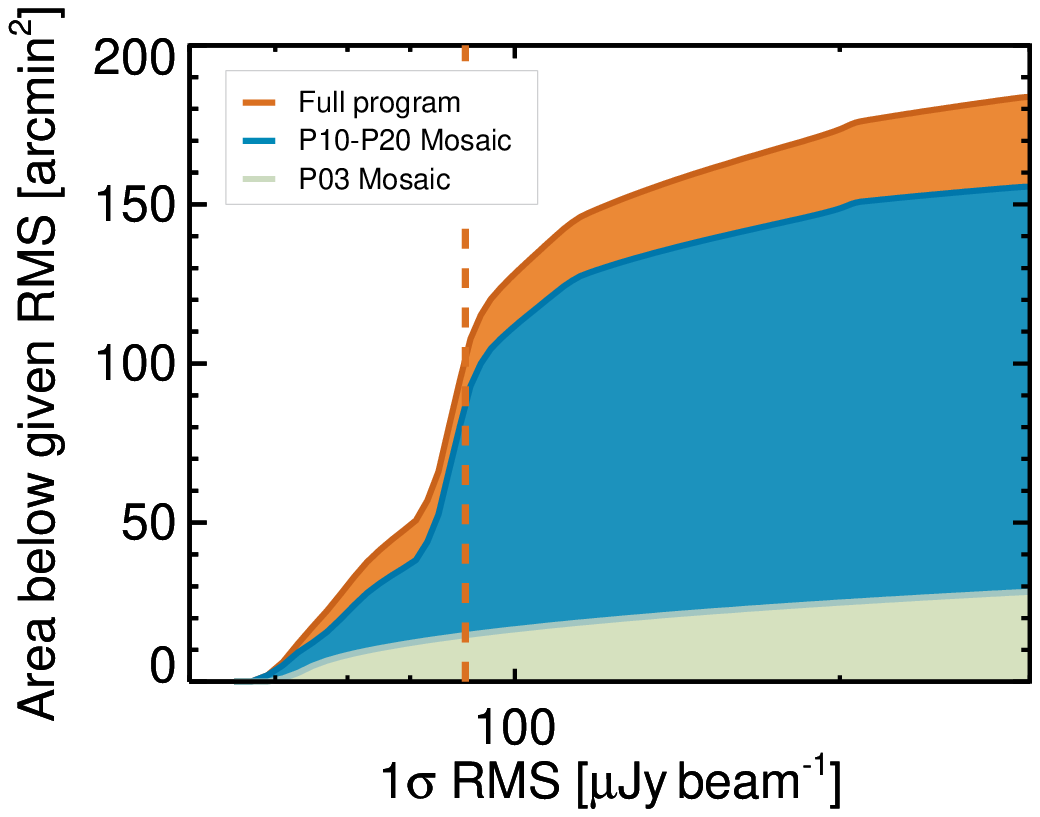}
\caption{The cumulative
  distribution of 1$\sigma$ survey depth and over what solid angle
  that RMS depth is achieved. The P03 mosaic alone (covering $28\,\rm arcmin^2$) is represented
  in pale green, while the P10--P20 mosaic (covering $156\,\rm arcmin^2$) is shown in blue. The proposed map depth ($90\,\mu$Jy) is illustrated by the orange dashed line. With a total area of $184\,\rm arcmin^2$ (orange line), the MORA program represents the  largest  ALMA  blank-field  survey  to-date  and  the only at 2 mm to search for dusty star-forming galaxies. 
\label{fig:area}}
\end{figure}

The MORA survey (ALMA project code: 2018.1.00231.S, PI: C. Casey) was originally designed to cover a contiguous area of  $\rm230\,$arcmin$^2$ in two different tunings in the 2\,mm band (centered at 147.3\,GHz and 139\,GHz, respectively). The project was only partially observed, covering $\rm184\,$arcmin$^2$ in two separate mosaics of  $\rm156\,$arcmin$^2$ and $\rm28\,$arcmin$^2$, respectively. 
The largest mosaic, which covers the positions 10 to 20 from the original design and so-called P10-P20, is centered at $\alpha\rm\approx10\,h\,00\,m\,17\,s$, $\delta\approx+02^\circ\,22'\,30''$ and covers an area of $6.6'\times23.6'$.
The smaller one (P03) is centered at $\alpha\rm\approx10\,h\,00\,m\,44\,s$, $\delta\approx+02^\circ\,22'\,30''$ and has $1.2'\times23.6'$ dimensions (see Figure \ref{fig:map}). The deepest portion of the mosaics has an RMS of $\sigma_{\rm 2mm}=60\,\mu\rm Jy\,beam$, with $101\,$arcmin$^2$ covered at or below the proposed map depth of $90\,\mu\rm Jy\,beam$ (see Figures \ref{fig:map} and \ref{fig:area}), with a typical beamsize of $\theta_{\rm FWHM}\approx1.8''\times1.4''$. This project represents the largest ALMA blank-field survey to-date and the only at 2\,mm to search for dusty star-forming galaxies.

\begin{figure}[t]\hspace{-0.2cm}
\includegraphics[width=0.51\textwidth]{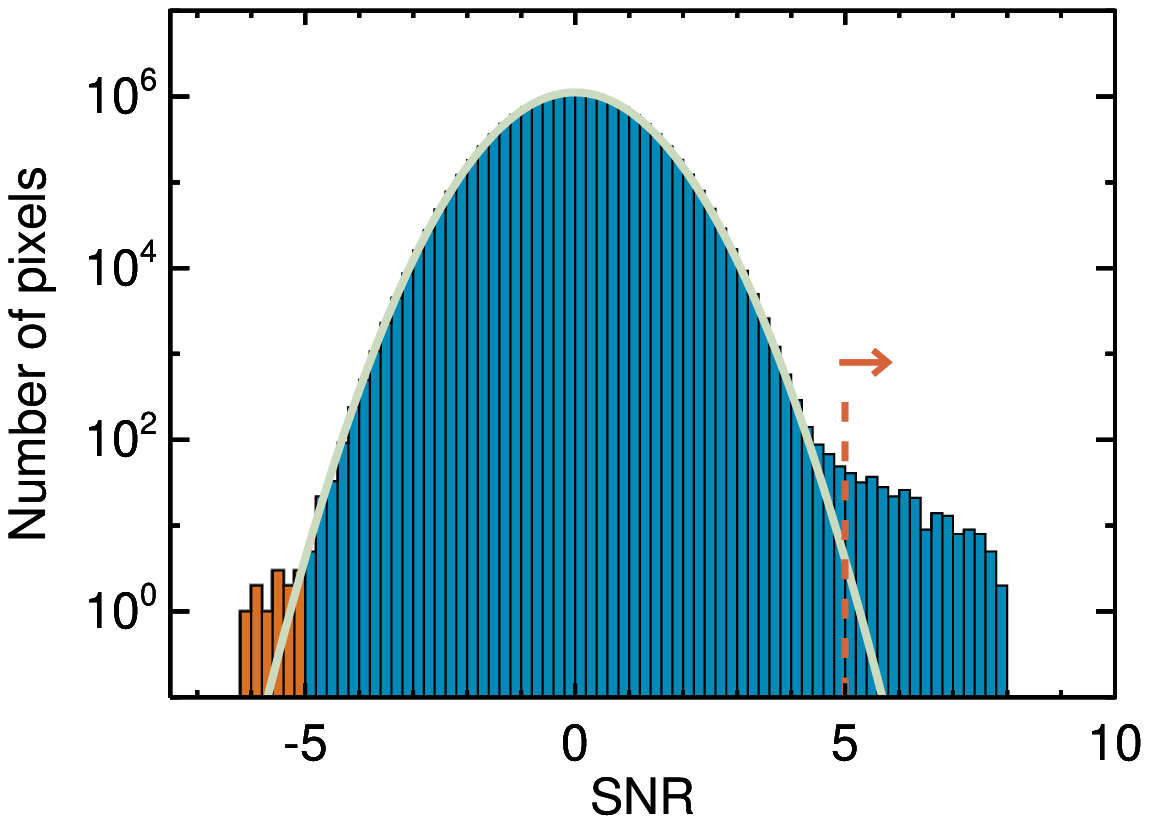}
\caption{Histogram of pixel values in the SNR maps. The distribution of pixel values within  $\rm -4.5\lesssim SNR\lesssim4.5$ is very well fitted by a Gaussian function (pale green solid line), which confirms the Gaussian properties of the map's noise. Above $\sim4.5\sigma$ the distribution starts to diverge due to the presence of positive sources. Our $5\sigma$ threshold to identify robust sources is indicated by the dashed orange line and arrow. All the pixels highlighted in orange, which have values below $-5\sigma$, lie within a region spanning approximately a beamsize. This negative signal could be associated with a single noise peak at $-6\sigma$ or with a detection of the SZ effect, as discussed in \S\ref{sec:MORA}.
\label{fig:pixels_histogram}}
\end{figure}

The signal-to-noise ratio (SNR) maps are well modelled by Gaussian statistics  within $-4.5\lesssim\rm SNR \lesssim 4.5$ (see Figure \ref{fig:pixels_histogram}). Above $4.5\sigma$ the excess of positive pixels comes from the detection of astronomical sources, as expected. Interestingly, there is also an excess of negative pixels with $\rm SNR<-5$. All these pixels (identified in orange in the figure) come from the same region and are associated with a single noise peak at $-6\sigma$ (plus the effect of beam smearing; see Figure \ref{fig:map}). Nevertheless, assuming Gaussian statistics, the probability of finding a $-6\sigma$ noise peak in our mapped area is $\lesssim0.5\%$. An alternative explanation might be the Sunayev-Zeldovich (SZ) effect (\citealt{Sunyaev1970a}) since its peak decrement is expected close to the observed frequency of our maps ($\sim145\,$GHz). A thorough discussion of the nature of this negative detection will be presented in a future work, after carrying out the required follow-up observations.

Casey et al. (in preparation) will present further details of the survey along with a description of the source catalog and other physical characteristics of the 2\,mm-selected galaxies.  Briefly, we first create a noise map using two different approaches. In the first one, we measure the standard deviation of all the pixels with primary beam responses between 0.95 and 1, which give us the minimum noise in our mosaic (or maximum depth); then we multiply
the inverse of the primary beam response map by this minimum noise vale, whose result is adopted as the noise map. In the second approach, we use a  2D boxcar-like function across the primary beam corrected flux mosaic (after applying a sigma clipping procedure to remove the bright sources) to estimate the noise in each pixel, measured as the standard deviation of the pixels within the squared region of the boxcar-like function. Both procedures give us consistent results within $\sim1$\%. Once we have a noise map, we simply divide the primary beam corrected flux map by this noise mosaic in order to obtain a SNR map (which is shown in Figure \ref{fig:map}). This SNR map is then used to find source candidates by searching for pixels above a SNR threshold.
Here, we use  the sources detected at $>5\sigma$ (where the false detection rate is expected to be $\lesssim8\%$; see \S\ref{secc:num_counts}) to calculate the first ALMA number counts at 2\,mm, as described below.

\subsection{The 2\,mm number counts}\label{secc:num_counts}

The cumulative number counts -- i.e. the number of galaxies above a certain flux density threshold, $S$, per unit area -- can be directly estimated by counting the number of detected galaxies as a function of flux density and making the required corrections for contamination, completeness and flux boosting.



Previous blind ALMA surveys have estimated a minimum, close-to-zero, contamination fraction due to false  detections (noise peaks) for SNR thresholds in the range of $\approx4-5$ (e.g. \citealt{Fujimoto2016a,Oteo2016a}). Nevertheless, despite adopting a conservative threshold of $5\sigma$ in our analysis (see Figure \ref{fig:pixels_histogram}), a thorough characterization of the contamination rate is required given the large area covered by the MORA survey and the large number of independent beams in the maps.


The false detection rate of the MORA survey is characterized as follows: first, we create noise maps with the same dimensions as the original maps under the assumption that the noise in the ALMA observations is well represented by Gaussian statistics (as has been shown before for other ALMA data, e.g. \citealt{Franco2018a}, and demonstrated in Figure \ref{fig:pixels_histogram}). Second, the maps are convolved with a beam representative of the average synthesized beam in our observations. Then, to characterize the contamination rate as a function of SNR, the map is scaled such that the final standard deviation of the whole distribution of pixels is equal to one. This is equivalent to a SNR map with no astronomical sources. Finally, we simply quantify the number of peaks above the adopted detection threshold using the same algorithm used to find the real sources. This process is repeated 100 times in order to measure the expected number of false sources and its corresponding uncertainty. At our adopted threshold of $5\sigma$, the expected number of false detections in the MORA catalog is $0.8\pm0.2$. This is in good agreement with the findings in Casey et al. in preparation, who found near-IR counterparts for all the MORA detections with the exception of the lowest significance source in the catalog.


To take into account this false detection rate in the number counts calculation, a statistical approach is adopted in which the contamination fraction is distributed among the 13 sources according to their SNRs with a proportionality given by a normal probability distribution (i.e. estimating the probability that a normal random variable is greater than the sources' SNRs). As expected,  this procedure gives a higher probability of false detection to the sources with the lowest SNRs and an (almost) negligible probability to those detected at $\gtrsim5.5\sigma$.




The survey completeness is calculated in a similar fashion in which artificial sources are inserted in the real maps (after removing the bright detections) followed by computing the ratio between the recovered sources and inserted sources. The artificial sources are inserted, one at a time, at random positions around the whole mosaic and then they are recovered with the same source extraction procedure used to build the real source catalog.
The survey completeness can be estimated as a function of input flux density, recovered flux density, detected SNR, local noise RMS, or a combination of these parameters.  A source is considered recovered if it is detected within 
a synthesized beam from the input position ($\sim1.5''$). In each flux density bin of $25\,\rm\mu Jy$, we repeat this process 10,000 times in order to sample the noise variation across the maps. 

The same set of simulations are used to
examine the flux-boosting effects (meaning sources' flux densities systematically biased upwards by noise and the presence of fainter undetected sources). This is done by estimating the average ratio between the sources' intrinsic flux density and the recovered flux density in the simulations. Again, this value depends on the flux density and on the local noise of each source (or similarly, SNR), and therefore, several flux boosting factors can be derived as a function of these parameters. In average,
we find flux boosting factors of around $5\%$ for those source detected just above our SNR threshold. These factors, which decrease with increasing SNR, are however
much lower than the uncertainties on the measured flux densities ($\sim20\%$ for a $5\sigma$ detection). 

The  number counts can then be directly estimated by counting the number of detected galaxies as a function of flux density and making the appropriate corrections for contamination, completeness and flux boosting. We estimate the contribution of a source 
with a deboosted flux density, $S_i$, and measured SNR, $\sigma_i$, to the cumulative number 
counts to be:
\begin{equation}
\eta_i(S_i,\sigma_i)=\frac{1-f_{\rm cont}(\sigma_i)}{\zeta(S_i)\,A_{\rm eff}},
\end{equation}
where $f_{\rm cont}(\sigma_i)$ is the estimated fraction of contamination at the measured SNR ($\sigma_i$) of the source, $\zeta$ is the corresponding completeness for a deboosted flux density $S_i$, and $A_{\rm eff}$ is the total area of the MORA maps used for source extraction (i.e. $184\,$sq.\,arcmin$^2$; see Figure \ref{fig:area}).
Finally, the  cumulative number counts, $N(>S)$, are estimated by the sum over all sources with a flux density higher than $S$, i.e. $N(>S)=\sum\limits_{i}\eta_i(S_i,\sigma_i)$. 

\begin{figure}[t]\hspace{-1.1cm}
\includegraphics[width=0.55\textwidth]{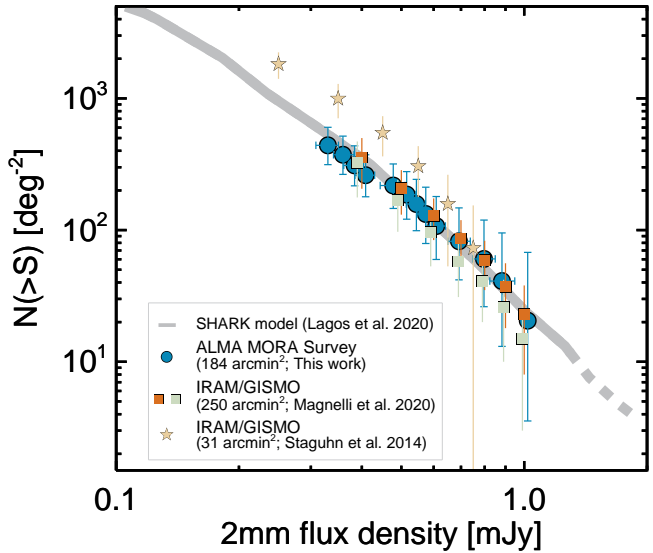}
\caption{The cumulative number counts derived from our main sample (SNR$>5\sigma$) are represented by the blue solid circles. Previous estimation of the number counts using the GISMO camera on the IRAM telescope are represented by the pale green and orange squares (\citealt{Magnelli2019a}) and the yellow stars (\citealt{Staguhn2014a}). Additionally, the predictions from the {\sc shark} galaxy evolution model (\citealt{Lagos2020}) are illustrated by the gray line. 
\label{fig:moras_num_counts}}
\end{figure}

\begin{deluxetable*}{cccccccc}[t]
\tablecaption{MORA survey 2\,mm number counts.  \label{table:moras_num_counts}}
\tablecolumns{4}
\tablenum{1}
\tablewidth{0pt}
\tablehead{
\multicolumn{4}{c}{Raw number counts} & \multicolumn{4}{c}{Star-forming galaxies number counts} \\
\colhead{$S_{\rm 2mm}$} &
\colhead{N($\rm >S_\nu$)} &
\colhead{$\delta N^-$} &
\colhead{$\delta N^+$ } &
\colhead{$S_{\rm 2mm}$} &
\colhead{N($\rm >S_\nu$)} &
\colhead{$\delta N^-$} &
\colhead{$\delta N^+$ } \\
\colhead{($\rm mJy$)} & 
\colhead{(deg$^{-2}$)} &
\colhead{ (deg$^{-2}$)} & 
\colhead{ (deg$^{-2}$)} &
\colhead{($\rm mJy$)} & 
\colhead{(deg$^{-2}$)} &
\colhead{ (deg$^{-2}$)} & 
\colhead{ (deg$^{-2}$)} \\
}
\startdata  
  0.33  &   440  &   120 &  60   &  0.33  &   380  &   110 &  150  \\
  0.36  &   370  &   110 &  140  &  0.36  &   320  &   100 &  130  \\
  0.39  &   310  &   90  &  130  &  0.39  &   270  &   80  &  110  \\
  0.41  &   260  &   80  &  110  &  0.48  &   220  &   70  &  100  \\
  0.48  &   220  &   70  &  100  &  0.52  &   185  &   64  &  91   \\
  0.52  &   185  &   64  &  91   &  0.54  &   157  &   59  &  85   \\
  0.54  &   157  &   59  &  85   &  0.57  &   131  &   53  &  79   \\
  0.57  &   131  &   53  &  79   &  0.61  &   106  &   47  &  72   \\
  0.61  &   106  &   47  &  72   &  0.68  &   83   &   41  &  65   \\
  0.68  &   83   &   41  &  65   &  0.78  &   60   &   34  &  59   \\
  0.78  &   60   &   34  &  59   &  0.88  &   41   &   27  &  53   \\
  0.88  &   41   &   27  &  53   &  1.03  &   21   &   17  &  47   \\
  1.03 &   21   &   17  &  47    &        &        &       &       \\
\enddata
\tablecomments{The star-forming galaxy number counts are estimated after removing source MORA-10, whose flux density falls below our detection threshold after removing the contribution from synchrotron emission.}
\end{deluxetable*}

In order to take into account the uncertainties associated with the correction factors (completeness and false detection) and flux densities in the estimation of the number counts, we perform a Monte Carlo simulation. In each realization, the adopted measured flux density  for each source, $S_i$, is extracted from a Gaussian distribution with a standard deviation 
equal to the measured error and centered at the observed value. Then, this new flux density is used to estimate a new SNR, $\sigma_i$, by dividing by the local noise. These SNRs are then used to update the contamination fraction and completeness accordingly, whose new values are drawn from Gaussian distributions. 
This procedure is repeated 100 times, with the mean values representing the final number counts and the 16th and 85th percentiles their associated confidence interval. Finally, given the relatively  small number of sources in our catalog, Poisson uncertainties are added in 
quadrature according to \citet{Gehrels1986a}. 
The MORA 2\,mm number counts derived in this work are reported in Table \ref{table:moras_num_counts} and are show in Figure \ref{fig:moras_num_counts}.

Given that the 2\,mm flux density of one of our sources, MORA-10, was found to be contaminated by synchrotron emission (at a $\sim40\%$ level; Casey et al. in preparation), we repeat all the process described above after removing this source, aiming at providing the 2\,mm number counts of star-forming galaxies only (note that the source would have been fallen below our detection threshold without the synchrotron emission).  This is particularly important since the model used in \S\ref{secc:model} does no take into account non-thermal emission, nevertheless, we highlight that the difference between the two estimations  is not significant (see Table \ref{table:moras_num_counts}).

At this wavelength, the only determinations of the number counts reported in the literature beyond our MORA survey measurements come from the GISMO/IRAM surveys reported in \cite{Staguhn2014a} and \cite{Magnelli2019a}, covering $31\,$arcmin$^2$ and $250\,$arcmin$^2$, respectively. The latter are in very good agreement with our estimations while the former lie above by a factor of $\approx2.0-2.5$. This over-estimation is thought to be caused by the uncertainties in the flux deboosting factors in the confusion-limited map of  \cite{Staguhn2014a}, as discussed in \cite{Magnelli2019a}, although  cosmic variance might be also important given the relatively small mapped area. The predictions from the {\sc shark} semi-analytic model of galaxy formation (\citealt{Lagos2020}) are also plotted in the figure, which are indeed in very good agreement with our estimations.

\section{Revising the 3\,mm number counts}\label{secc:3mm_num_counts}

\cite{Zavala2018c} compiled archival ALMA band 3 observations toward three extragalactic legacy fields: COSMOS, CDF-S, and the UDS, resulting in a total of 135 individual maps adding up a total area of $\approx200\,\rm arcmin^2$. 
After masking out the original targets of these observations and any galaxies that were potentially physically associated with them, they  derived the first ALMA galaxy number counts at $3\,$mm using a total of 13 sources detected above $5\sigma$. 

Our recent ALMA 2mm + 3mm follow-up observations on this sample (ALMA projects: 2018.1.00478.S and 2019.1.00838.S; PI: J. Zavala) show that three of these sources (ALMA-3mm.14, ALMA-3mm.15, and ALMA-3mm.16) are not recovered, meaning they are likely spurious detections. Two of these sources were indeed noted to have a spectral index which might be inconsistent with thermal emission in \cite{Zavala2018c}, nevertheless, without further data at the time, the low SNRs ($5.0-5.2$) prevented a firm conclusion and therefore they were included in the previous number counts estimation. 

Thus, we have revised the false detection rate to be higher than the value reported in the original work of \cite{Zavala2018c}, for which only one source was expected to be false.  This highlights the complexity of the interferometric data, particularly when using observations with different beamsizes, integration times, and array configurations, as is typical for datasets derived from archival projects.

\begin{deluxetable}{cccc}[h]
\tablecaption{Revised ALMA archival 3\,mm number counts.  \label{table:revised_num_counts}}
\tablecolumns{4}
\tablenum{2}
\tablewidth{0pt}
\tablehead{
\colhead{$S_{\rm 3mm}$} &
\colhead{N($\rm >S_\nu$)} &
\colhead{$\delta N^-$} &
\colhead{$\delta N^+$ } \\ 
\colhead{($\rm \mu Jy$)} & 
\colhead{(deg$^{-2}$)} &
\colhead{ (deg$^{-2}$)} & 
\colhead{ (deg$^{-2}$)}
}
\startdata 33   &  4440 &  1200 &   1990  \\
52   &  2140 &  640  &   1000  \\
63   &  1420 &  470  &   710   \\
79   &  990  &  360  &   540   \\
98   &  750  &  290  &   450   \\
109  &  550  &  230  &   370   \\
117  &  410  &  190  &   320   \\
127  &  290  &  150  &   280   \\
137  &  170  &  110  &   230   \\
167  &  74   &  55   &   170   \\
\enddata
\end{deluxetable}

Here, we revised the 3\,mm number counts following exactly the same procedure as described in \cite{Zavala2018c} but removing the three spurious sources. Additionally, we updated the flux density of the source ALMA-3mm.03 (a.k.a. ASPECS-3mm.1) since the reported value in \cite{Zavala2018c} seems to be contaminated by a bright CO emission line (\citealt{Gonzalez-Lopez2019a}); we thus adopt the flux density reported in the ASPECS catalog.

The revised 3\,mm number counts are reported in Table \ref{table:revised_num_counts} and shown in Figure \ref{fig:num_counts}. The new values are significantly different at the bright end ($S_{\rm 3mm}>0.2\,$mJy) since two of the false detections were the brightest galaxies in the sample\footnote{Given the inhomogeneous observations in the archival data, a low signal-to-noise ratio does not necessary imply a low flux density since each map has a different noise r.m.s. depth. The two brightest false detections were indeed detected in areas with large r.m.s values.}. At fainter flux densities, the updated number counts are a factor of $\sim1.5\times$ lower, and thus, in better agreement with the ASPECS results (\citealt{Gonzalez-Lopez2019a}), although still a factor of $\sim2\times$ higher. Nevertheless, the two number counts are  consistent with each other within the uncertainties. The difference is thus not statistically significant.

\section{Constraining the IR luminosity function and dust-obscured star formation rate density}\label{secc4}

In this section we use a backward evolution model in combination with the state-of-the-art FIR/mm surveys to constrain the IR luminosity function.
The model draws spectral energy distributions (SEDs) from the known breadth and characteristics of dusty galaxies, then works backward to discriminate between different luminosity function scenarios using the galaxy number counts as constraints. 
Finally, the dust-obscured star formation rate density is estimated by integrating the best-fit IR luminosity function (see \citealt{Zavala2018c} for a similar analysis).

\subsection{Model description}\label{secc:model}
The adopted backwards evolution model is described in detail in \cite{Casey2018b,Casey2018a}, where the specifics of the model and the assumed values  for each parameter can be found. A summary of the salient characteristics and assumptions follows.

The model combines a parameterized evolving galaxy IR luminosity
function  with the thermal SED properties of galaxies' dust emission to make predictions for galaxy (sub-)millimeter surveys. 

Galaxies' SEDs are described by a modified black-body function of the form  $S_\nu\propto(1-e^{-(\nu/\nu_0)^{\beta_{\rm em}}})B_\nu(T_{\rm dust})$ with an extra mid-infrared power law to account for hotter dust emission at $T\gg T_{\rm dust}$ (e.g. \citealt{Casey2012a}). The SED library  follows the well-known luminosity-temperature relationship in terms of $L_{\rm IR}-\lambda_{\rm peak}$ and takes into account the observed dispersion in this relation. We highlight that using the peak wavelength rather than  dust temperature minimizes the impact of the assumed effective dust opacity in the model ($\tau=1$ at $\lambda_{\rm rest}=100\,\rm\mu m$), which could  change the derived dust temperature but has a minor impact on the measured $\lambda_{\rm peak}$. Note that although this relation is not assumed to evolve with redshift (which is supported by the current data at least up to $z\sim4.5$ \citealt{Casey2018a,Dudzeviciute2020a}), if one selects galaxies on the main sequence, a redshift evolution of $\lambda_{\rm peak}$ (proportionally to dust temperature) naturally arises. This evolution, which is driven by the underlying evolution of the main sequence of star-forming galaxies, is in line with recent results from the literature (e.g.  \citealt{Bethermin2015a,Schreiber2015a,Magdis2017a}).

As is common practice in the literature (e.g. \citealt{Sanders2003a,Magnelli2011a,Magnelli2013a,Lim2020a}), the assumed IR luminosity function in the model is described by a double power law of the form:
\begin{equation}
\Phi(L,z) =
  \begin{cases} 
      \Phi_\star(z)\left(\frac{L}{L_\star(z)}\right)^{\alpha_{\rm LF}},               & \mbox{if } L<L_\star, \\
      \Phi_\star(z)\left(\frac{L}{L_\star(z)} \right)^{\beta_{\rm LF}},               & \mbox{if } L\ge L_\star,
  \end{cases}
\end{equation}
where $\alpha_{\rm LF}$ and $\beta_{\rm LF}$, represent the slopes at faint and bright luminosities, respectively, while $L_\star$ and $\Phi_\star$ are the characteristic galaxy luminosity and characteristic number density, two fundamental parameters that are allowed to evolve with redshift. 

Following previous works from the literature, the characteristic number density  is assumed to evolve as:
\begin{equation}
\Phi_\star \propto
  \begin{cases} 
       (1+z)^{\Psi_1},  & \mbox{if } z<z_{\rm turn}, \\
     (1+z)^{\Psi_2},   & \mbox{if } z\ge z_{\rm turn}.
  \end{cases}
\end{equation}

Significant observational efforts have provided good constraints on the evolution of $\Phi_\star$ at low redshifts, revealing a flat trend with $\Psi_1$ close to zero  and a turnover redshift of $z_{\rm turn}\approx2$, in-line with the peak of the cosmic star formation rate density (CSFRD;  \citealt{Magnelli2011a,Gruppioni2013a,Lim2020a}). Although it is clear that at higher redshifts the evolution is significantly steeper, the current estimations are highly uncertain and limited to $z\lesssim4$, with values ranging  from $\Psi_2\sim-6$ (or lower; \citealt{Koprowski2017a}) to $-2$ (or even higher; \citealt{Rowan-Robinson2016a}), with a few measurements in between. Indeed,  \citet{Zavala2018c} found an intermediate value of  $\Psi_2\approx-4.2$ using the initial 3\,mm sample.

In this work, we aim to constrain two of the most important unknowns of the galaxy luminosity function: the faint-end slope, $\alpha_{\rm LF}$, and the redshift evolution of the characteristic number density at high redshift driven by $\Psi_2$ (since $\Phi_\star\propto(1+z)^{\Psi_2}$ for $z\gtrsim2$).  We also explore different values of the dust emissivity index, $\beta_{\rm em}$, in order to minimize any possible bias and to test possible degeneracies between the different parameters.

For the rest of the model parameters, we adopt the same values as in \citet{Casey2018a} and \citet{Zavala2018c}. We refer the reader to \citeauthor{Casey2018a} (\citeyear{Casey2018a}, and particularly to its appendix A.1) for a detail discussion about these choices and their impact on the model. In short, the bright-end slope of the luminosity function is fixed to be $\beta_{\rm LF}=-3.0$. This value is relatively well-constrained (at least up to $z\sim4$) with a measured dispersion of $1\sigma\approx0.15$.
The evolution of the characteristic luminosity, $L_\star$, follows $L_\star\propto(1+z)^{\gamma_1}$ for $z<z_{\rm turn}$ and $L_\star\propto(1+z)^{\gamma_2}$ for $z\ge z_{\rm turn}$. While the evolution at lower redshifts ($z<z_{\rm turn}$) is well-characterized (we adopt $\gamma_1=2.8$), the evolution of $L_\star$ beyond $z_{\rm turn}$ is more uncertain. The model adopts $\gamma_2=1.0$, implying that $L_\star$ continues to evolve upward toward higher redshifts, in line with expectations for hierarchical structure formation and cosmic downsizing. Additionally, adopting $\gamma_2=1.0$ makes the $L_\star$ evolution consistent with the $L_\star$ of the quasar luminosity function (\citealt{Hopkins2007a}). Adopting a reversal evolution ($\gamma_2\le0$) under-predicts the number counts at wavelengths longer than $\sim850\rm\,\mu m$, regardless of the adopted number density (for any $\Psi_2<-1.5$), while a more rapid evolution with $\gamma_2\ge1.5$ would imply very bright values for $L_\star$ at  $z\sim5$, which might even exceed the luminosity of the brightest DSGFs known to-date.  Moreover, in order to match the contribution from LIRGs and ULIRGs to the cosmic SFRD at $z<2.5$ to those reported in the literature, a modest dependence of $z_{\rm turn}$ on $\Psi_2$ of the form $z_{\rm turn}=1.6-0.09\Psi_2$ has to be adopted (see \citealt{Casey2018a} for details).

This model has some caveats that the reader should keep in mind. 
The exact values of $L_0$ and $\Phi_0$ are correlated with $\gamma_1$ and $\Psi_1$, respectively\footnote{We refer the reader to equations 8 and 9 from \citet{Casey2018a} for the exact parametrization of the evolution of $L_\star$ and $\Phi_\star$ and the role of $L_0$ and $\Phi_0$.}. As discussed in detailed by \citet{Casey2018a}. The adopted values ($L_0=10^{11.1}\,L_\odot$, $\Phi_o=10^{-3.5}\rm\,Mpc^{-3}\,dex^{-1}$, $\gamma_1=2.8$, and $\Psi_1=0$) were chosen to simultaneously reproduce: {\it(i)} the IRLF at $z\lesssim2$;  {\it(ii)} the reported values of $L_\star$ and $\Phi_\star$ from the literature; and  {\it(iii)} the relative contribution from LIRGs and ULIRGs to the total SFRD within $0<z<2$.  This combination of values are, however, not necessarily unique. Another caveat that the reader should keep in mind is the assumption of a  fixed emissivity spectral index at all redshifts, and although several values were explored in the analysis (from $\beta=1.6$ to 2.6, see Figure \ref{fig:triangle_plot}), each realization adopts a single value across all redshifts for all sources.  This is not necessarily in disagreement with previous literature, given the current limited data. Although a spread on $\beta$ has been found across different galaxies, these variations have not yet been shown to correlate with redshift or other galaxy characteristics that may impact the model. Indeed, there are clear examples of $\beta\sim1.5$ to $\sim2.0$ in both local galaxies (e.g. \citealt{Remy-Ruyer2013a}) and high-redshift systems (e.g. \citealt{Jin2019a}). 
The same is true for the adopted slopes of the luminosity function at both the faint and bright ends ($\alpha_{\rm LF}$ and $\beta_{\rm LF}$, respectively). Although a redshift evolution is possible given that we know similar parameters do evolve for the UV luminosity function (e.g. \citealt{Finkelstein2015a}), the lack of samples of galaxies with spectroscopic redshifts covering a wide range of IR luminosities prevents its confirmation (or rejection). Indeed, non-evolving slopes have been been adopted in several works in the literature (e.g. \citealt{Magnelli2011a,Magnelli2013a,Gruppioni2013a,Lim2020a}).  Therefore, in the absence of more information, we choose to fix these quantities. The impact of these assumptions on our results are further discussed below.

\subsection{Fitting methodology and data constraints}

To find the best-fit model parameters,
first, the different evolutionary scenarios of the IRLF are combined with galaxies' SEDs to create mock observations that resemble the real surveys in terms of wavelength, noise depth, and angular resolution, with the CMB heating effects taken into account (following \citealt{daCunha2013a}). Second, the respective  number counts are estimated in a similar fashion as in the real observational works. 
These simulated number counts are then compared to the measured number counts in a joint analysis that combines multiple observations at different wavelengths simultaneously. Finally, the best-fit model parameters that better reproduce all the observed number counts are derived.

The cumulative number counts are preferred over the differential number counts since the latter suffer from larger uncertainties due to the relative small number of sources in each flux density bin. Though we test that adopting the differential number counts instead do not change significantly our results.

As discussed in \cite{Casey2018b,Casey2018a},
observations at long-wavelengths ($\lambda\gtrsim1\,$mm) are ideal to distinguish between the different scenarios of the model since the majority of constraining data sets at shorter wavelengths only inform about the evolution of the IRLF at $z\lesssim3.0$, where the model parameters are already relatively well understood. Therefore, here we only use number counts at 1.2\,mm, 2\,mm and 3\,mm as constraints, which also reduces significantly the computational cost of the fitting analysis, which would be prohibitive otherwise. We adopt the MORA 2\,mm number counts described above (see \S\ref{secc:num_counts}), the updated 3\,mm number counts reported in this work (see \S\ref{secc:3mm_num_counts}), and the 1.2\,mm and 3\,mm number counts from the ASPECS survey (\citealt{Gonzalez-Lopez2019a,Gonzalez-Lopez2020a}). Altogether, these surveys add a total of $\sim400\,$sq.\,arcmin surveyed area and around 60 detected galaxies. Note that AGN contribution or non-thermal emission (e.g. synchrotron emission) is expected to be negligible at these wavebands and at the flux density ranges explored here\footnote{From all the sources used in this work, we found a non-negligible contamination from synchrotron emission only in one source, namely MORA-10. This source has been removed from our analysis (see \S\ref{secc:num_counts}).} (see discussion in \citealt{Casey2018b,Zavala2018c}).

Once the fiducial number counts have been defined,  to derive the best-fit model parameters that better reproduce these  number counts we use two different methods: a maximum likelihood approach and a multi-dimensional minimization algorithm.

The maximum likelihood approach is done in a similar fashion as reported by  \cite{Zavala2018c}. We create a three-dimensional grid, representing the parameter space formed by
$\alpha_{\rm LF}$, $\Psi_2$, and $\beta_{\rm em}$, with values ranging from $\alpha_{\rm LF}: [-0.6,-0.2]$ in steps of 0.05, $\Psi_2: [-8.8,-2.2]$ in steps of 0.4, and $\beta_{\rm em}: [1.5,2.6]$ in steps of 0.05.
For each combination of the parameters subset (corresponding to different points in the grid), the likelihood function of the measured cumulative number counts relative to the simulated number counts is estimated. Finally, 
the best-fit model is assumed to be the one with the highest probability, and the confidence intervals at  68\%, 95\%, and 99.7\% are
obtained  by integrating the normalized likelihood distribution. 



\begin{figure*}[ht]\hspace{-0.3cm}
\includegraphics[width=1.03\textwidth]{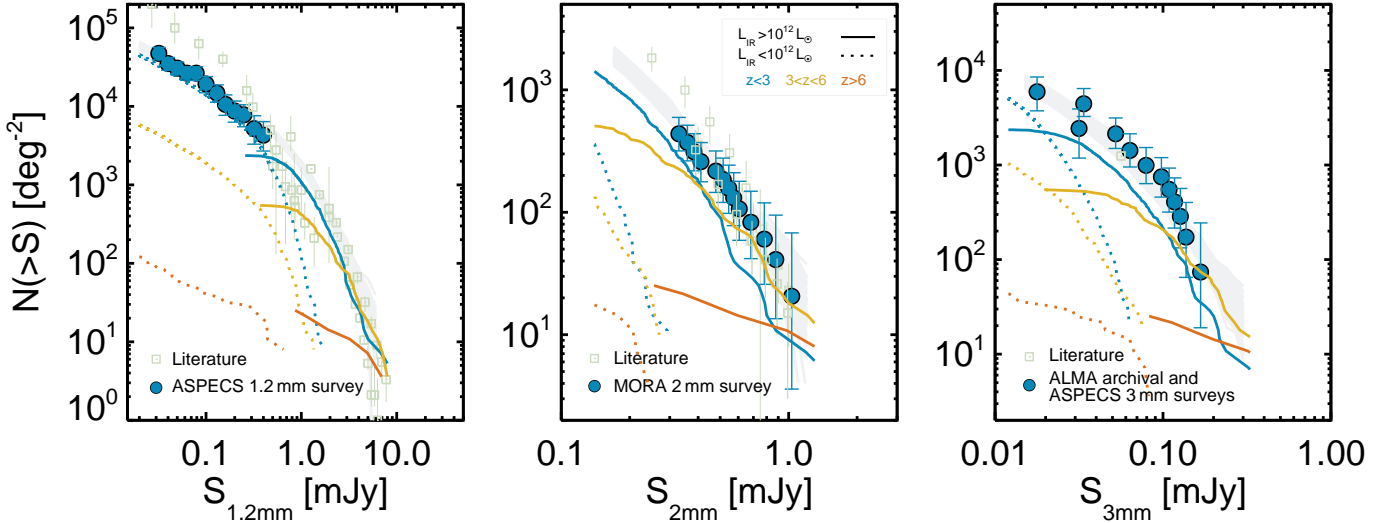}\vspace{0.7cm}
\caption{Cumulative number counts -- galaxies above a given flux density per unit area -- at 1.2\,mm (left panel), 2\,mm (middle panel), and 3\,mm (right panel). The data points used for the fitting analysis are illustrated as blue points 
    while other measurements from the literature are plotted as green squares. The best-fit number counts from the model, represented by the gray lines, nicely reproduce the number counts at the three different wavelengths simultaneously, including those not used in the fitting procedure   as well as number counts at shorter wavelengths (see Appendix \ref{appendix}), spanning more than a decade in wavelength and order of magnitudes in flux density. The best-fit number counts from the model are also broken down into two ranges of luminosity and three redshift bins.  Galaxies with $L_{\rm IR}>10^{12}\,L_\odot$ are shown in solid lines while those with $L_{\rm IR}<10^{12}\,L_\odot$ are shown in dotted lines. Similarly, the three different redshift ranges are illustrated by different colors: blue for $z<3$, gold for $3<z<6$, and orange for $z>$6.
\label{fig:num_counts}}
\end{figure*}

The Nelder–Mead optimization method (a.k.a the Amoeba method; \citealt{Nelder1965a,numerical_recipes}) relies on a downhill simplex algorithm to perform a multidimensional minimization of a given parameter, in this case, the square differences ($\chi^2$) between the real and the modeled cumulative number counts. In each realization, we randomly vary the initial starting point of the search, with each search limited to 100 evaluations. During this analysis, we introduce a fourth parameter, $z_{\rm cutoff}$, following \cite{Zavala2018c}, which represents the redshift above which no more dusty galaxies exist (with a range of  $z_{\rm cutoff}=5.5-9.0$). 
Nevertheless, we find that $z_{\rm cutoff}$ is not well constrained by the data and has a minor impact on the analysis.




As shown in Figure \ref{fig:num_counts}, the best-fit number counts recovered from the model are in good agreement with the measured values, reproducing simultaneously not only the data used in the fitting but also the rest of measurements reported in the literature, as discussed below.

At 1.2\,mm, beyond the ASPECS number counts, the model predictions nicely reproduce the number counts from the 1.6\,deg$^2$ AzTEC surveys  reported in \cite{Scott2012a} and those from the ALMA follow-up survey of SCUBA-2 galaxies detected over $\sim1\rm\,deg^2$ reported by \cite{Stach2018a}\footnote{The results from  AzTEC observations and other ALMA surveys at 1.1\,mm were scaled  by a factor of 0.8 while those from the ALMA/SCUBA-2 survey at $850\,\rm\mu m$ were scaled by a factor of 0.4.}, which probe the brightest flux densities. At fainter flux densities, the model also reproduces the more recent ALMA results from \cite{Dunlop2017a} and \cite{Umehata2017a}. The number counts from the GOODS-ALMA survey (\citealt{Franco2018a}; covering the flux density range of $S_{\rm1.2mm}\approx0.5-2\,$mJy) are, however, lower than our estimations, and those reported by 
\citet{Fujimoto2016a} (which probe the faintest flux densities in the figure) lie above our estimations. 
Note, however,  that the measurements reported by \cite{Fujimoto2016a} come from gravitationally lensed fields, and therefore, these discrepancies might be caused by the uncertainties in the magnification factors and  small survey area  (see discussion in \citealt{Gonzalez-Lopez2020a}). The discrepancy with the \citet{Franco2018a} results is likely related to the relatively low completeness associated with their high angular resolution observations ($\theta_{\rm FWHM}=0.29''$ and $\theta_{\rm FWHM}=0.60''$ after filtering the data).

At 2\,mm, the only determinations of the number counts reported in the literature beyond our MORA survey measurements come from the GISMO/IRAM surveys reported in \cite{Staguhn2014a} and \cite{Magnelli2019a}, covering $31\,$arcmin$^2$ and $250\,$arcmin$^2$, respectively. The latter are in very good agreement with the model predictions while the former lie above by a factor of $\approx2.0-2.5$. This difference is thought to be caused by the uncertainties in the flux deboosting factors in the confusion-limited map of  \cite{Staguhn2014a}, as discussed in \cite{Magnelli2019a}, plus the possible effects of cosmic variance (see \S\ref{secc:num_counts}).

At 3\,mm, the only measurement plotted in Figure \ref{fig:num_counts} which was not used in our analysis, is the brightest bin of the ASPECS survey, which represents only an upper limit. This value is, however, in good agreement with the model predictions. 

The success of the model can also be illustrated by its predicting power at other wavelengths.
The model does reproduce the number counts at shorter wavelengths, from $70\,\mu\rm m$ to $850\,\mu\rm m$ spanning over two decades of observations over hundreds of square degrees (see Appendix \ref{appendix}).

 \subsection{The IRLF and best-fit parameters}\label{secc:lum_func}

\begin{figure*}[ht]
\includegraphics[width=\textwidth]{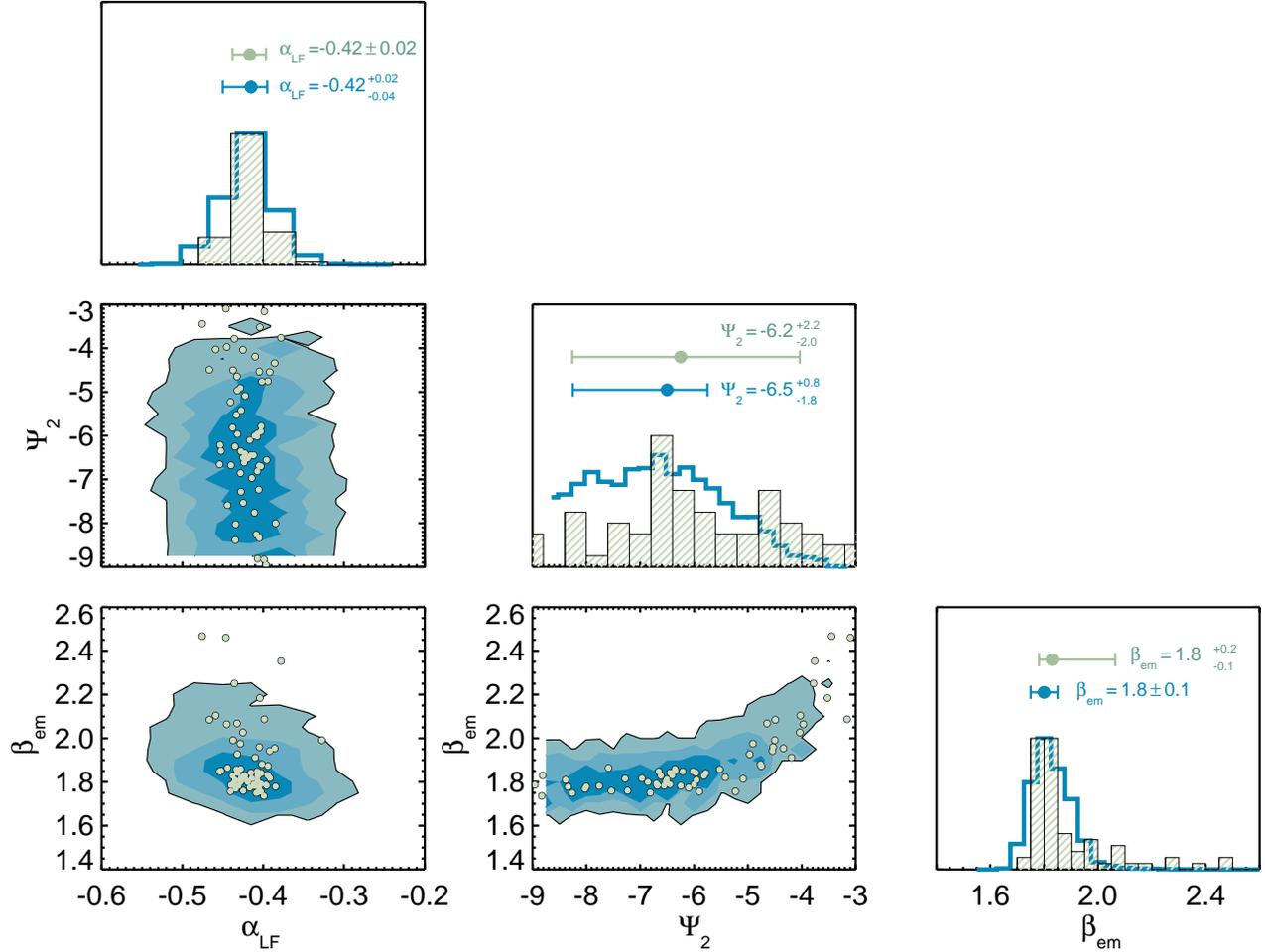}
\caption{Constraints on the IRLF model parameters  $\alpha_{\rm LF}$ (the faint-end slope of the luminosity function), $\Psi_2$ (which governs the evolution of $\Phi_\star$ at high redshift given $\Phi_\star\propto(1+z)^{\Psi_2}$), and $\beta_{\rm em}$ (the dust emissivity index). The confidence regions at the 68\%, 95\%, and 99.7\% for each parameter derived using the maximum likelihood approach are represented by the blue contours (from dark blue to light blue).
    Additionally,  the best-fit values extracted from 100 realizations of the Nelder–Mead multidimensional minimization algorithm (amoeba) are illustrated with the green solid circles. On the top of each column, we show the respective 1D marginalized probability distribution derived from the maximum likelihood approach (solid blue line), and the histogram of best-fit values found by the Nelder–Mead method (green dashed histograms).
    The best-fit values derived from both methods and their corresponding $1\sigma$ uncertainties (68\% C.I.) are shown on the top of each row. Consistent results are derived from the two different approaches.
\label{fig:triangle_plot}}
\end{figure*}

As it is clearly shown in Figure \ref{fig:triangle_plot}, both methods provide consistent results and good constraints on the model parameters governing the IRLF while successfully reproducing the observed number counts (see Figures \ref{fig:num_counts} and \ref{fig:num_counts_short_waves}).

The faint-end slope of the infrared luminosity function is found to be flat, with  $\alpha_{\rm LF}=-0.42^{+0.02}_{-0.04}$. This is in line with recent studies based on the deepest single-dish observations and interferometric surveys (although those were limited to lower redshifts), with reported values of $\alpha_{\rm LF}=-0.4$ (\citealt{Koprowski2017a}) and  $-0.5\pm0.7$  (\citealt{Lim2020a}), and with the observed flattening in the 1.1\,mm luminosity function (\citealt{Popping2020a}).

Although the faint-end slope is fixed for all redshifts in our model, we highlight that an evolution to steeper values with increasing redshift is inconsistent with our data and with the very low number of $z>4$ galaxies detected in the deepest ALMA surveys (\citealt{Dunlop2017a,Hatsukade2018a,Aravena2020a}). Indeed, most of the sources in the ASPECS survey lie at $z<3$ (\citealt{Aravena2020a}), in good agreement with our model predictions (see Figure \ref{fig:num_counts}).
This IR flat slope contrasts with the steep faint-end slope of the UV luminosity function at high redshift, which ranges from $-1.6$ at $z\sim4$ to $-2.0$ at $z\sim7$ (e.g. \citealt{McLure2013a,Finkelstein2015a}).
As a consequence, deep pencil-beam observations at (sub-)mm wavelengths would not significantly increase the number of detected sources, as is commonly the case in the rest-frame UV/optical observations (e.g. \citealt{Ferguson2000a}).

Regarding the evolution of the characteristic number density at high redshift, the best-fit $\Psi_2$ value of $-6.5^{+0.8}_{-1.8}$ implies a steep redshift evolution of the IR number density ($\Phi_\star\propto(1+z)^{-6.5^{+0.8}_{-1.8}}$), which disfavours the dust-rich hypothetical model discussed in \cite{Casey2018a}, for which a value of  $\Phi_\star\propto(1+z)^{-2.5}$ was adopted. This indicates  that dusty star-forming galaxies are indeed {\it rare} at early epochs.  This steep evolution is also in line with the rapid drop-off of the quasar luminosity function at high redshift (\citealt{Hopkins2007a}) and similar to the number density evolution of UV-bright (M$_{\rm UV}=-$21) galaxies ($\propto(1+z)^{-5.9}$; \citealt{Finkelstein2015a}), which might suggest that these galaxies occupy similar dark matter halos. 

There is, though, a caveat related to the effective dust optical opacity of our SEDs that the reader should keep in mind. Although, as mentioned before, the SEDs are parameterized in term of $\lambda_{\rm peak}$ rather than dust temperature, incorporating the CMB effects on the heating and detectability of the sources requires an estimation of the  dust temperature (\citealt{daCunha2013a}). Assuming an optically thin opacity form would result in lower dust temperatures for the same $\lambda_{\rm peak}$ than those derived from the optically thick model, decreasing the contrast between the dust emission and the CMB. This implies that, in order to match the measured number counts, a higher number of galaxies would be required compared to the optically thick model, i.e. a higher $\Psi_2$ value (see discussion in \citealt{Zavala2018c}). Nevertheless,
this effect is only important if the number counts are dominated by very high-redshift ($z>6$) sources, which our results suggest is likely not the case.

Finally, Figure \ref{fig:triangle_plot} shows that the dust emissivity spectral index is well-constrained to be $\beta_{\rm em}=1.8\pm0.1$, in very good agreement with the values reported in the literature (e.g. \citealt{Dunne2011a,Galliano2018a}; although we note that measurements of this parameter
at high redshifts are scarce). This figure also reveals, unsurprisingly, a mild correlation between
$\beta_{\rm em}$ and $\Psi_2$ since higher values of $\beta_{\rm em}$  imply  lower flux densities at long wavelengths (for a given
IR luminosity, redshift, and dust temperature).

We highlight that despite the large range of values explored for the three different parameters and despite the caveats described above, $\alpha_{\rm LF}$ and   $\beta_{\rm em}$ are  well-constrained, and while the uncertainties on $\Psi_2$ seem large, our constrains rule out values grater than $\Psi_2\sim-4$, which would imply a significantly larger number of DSFGs at high-redshifts.

\subsection{The history of dust-obscured star formation}

\begin{figure*}[ht]
\centering
\includegraphics[width=0.62\textwidth]{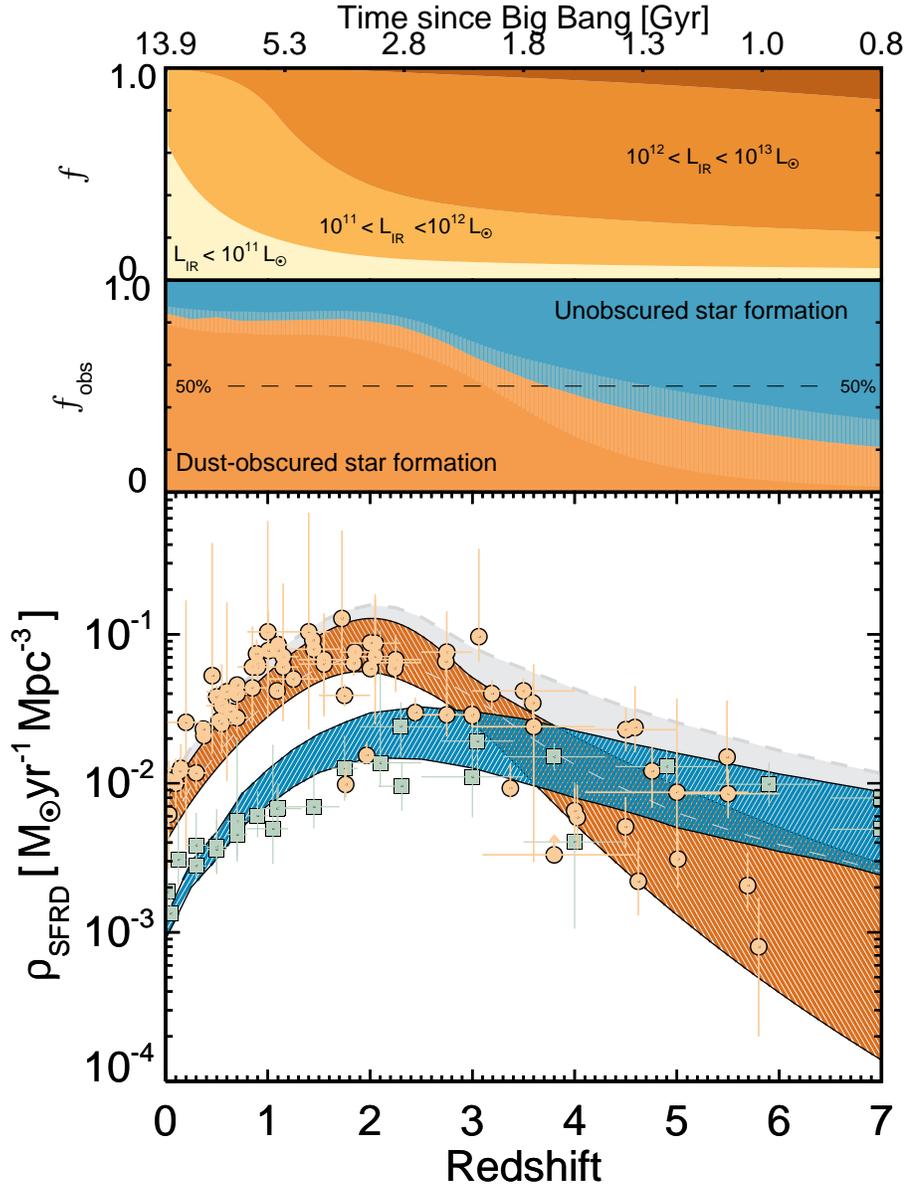}
\caption{The inferred dust-obscured star formation history is illustrated by the orange shaded region in the bottom panel. For comparison, we  plot independent measurements from the literature based on IR/sub-mm  and UV surveys (orange circles and blue squares, respectively) and the average unobscured star-formation derived from rest-frame UV optical surveys (i.e. not corrected for dust attenuation; blue shaded region; \citealt{Finkelstein2015a}). The total inferred SFRD derived in this work is shown in gray. The uncertainties in our estimation include those from the best-fit parameters and cosmic variance.  The middle panel represents the  fraction of obscured star formation, $\rm SF_{obs}/(SF_{obs}+SF_{unobs})$, and its associated uncertainty (lighter shaded area). The contribution of dust-obscured galaxies, which dominates the cosmic star-formation history through the last $\sim12\,$Gyr, rapidly decreases beyond its maximum, reaching values that are comparable to the unobscured star formation traced by the rest-frame UV/optical surveys by $z\approx4-5$. 
The top panel represents the contribution from galaxies with different luminosity ranges to the dust-obscured SFRD, being dominated by ULIRGs (ultra-luminous infrared galaxies; $10^{12}<L_{\rm IR}<10^{13}\,L_\odot$) and  LIRGs ($10^{11}<L_{\rm IR}<10^{12}\,L_\odot$). 
\label{fig:sfrd}}
\end{figure*}

The star formation rate density as a function of redshift can now be calculated by integrating the best-fit infrared galaxy luminosity function\footnote{The IR luminosity function is integrated over the interval log$(L/L_\odot):[9,13.8]$. A change of these limits has a minor impact on our results since the majority of the contribution arise from galaxies with  luminosities in the range of $L_{\rm IR}=10^{11}-10^{13}\,L_\odot$ (see Figure \ref{fig:sfrd}).} and propagating the associated uncertainties.

An important source of uncertainty is the field-to-field variation due to the large scale structure of the Universe, which is known as cosmic variance. To infer its impact in our estimations, we adopt the results from a model for the dust continuum number counts of galaxies (\citealt{Popping2020a}) that builds upon the {\sc UniverseMachine} model (\citealt{Behroozi2019a}), from which 1000 different light cones of 400\,sq.\,arcmin (approximately the combined area the MORA, ASPECS, and the 3\,mm archival surveys) are used to measure the variance in the number of detected sources. The cosmic variance is then defined following \citet{Moster2011a}:
\begin{equation}
\sigma^2 \equiv \frac{\langle N^2\rangle - \langle N\rangle^2}{\langle N\rangle^2} - \frac{1}{\langle N\rangle},
\end{equation}
where $\langle N^2\rangle$ and $\langle N\rangle$ are the  variance and mean number of sources in the light cones.
At the most critical redshifts of our work ($z>3$), the cosmic variance is estimated to be $\approx35\%$.  At $z\approx7.0$, the cosmic variance increases significantly and our data-sets  suffer a loss of constraining power. Furthermore, an increasing fraction of the star formation activity is expected to be in highly clustered structures (i.e. galaxy proto-clusters; \citealt{Chiang2017a}) that might have been missed in our surveys. Therefore, the model predictions presented in this work are limited to $z\lesssim7.0$.

The inferred dust-obscured star-formation history and its associated uncertainty, including that from cosmic variance, is presented in Figure \ref{fig:sfrd}. 
As can be seen, it is dominated by galaxies with $L_{\rm IR}\approx10^{11}-10^{13}$, a population that our surveys are particularly sensitive to (see Figure \ref{fig:lum_limits} in Appendix \ref{appendix2}).
Figure \ref{fig:sfrd} also includes other determinations of the CSFRD from the literature, which are split in two groups: FIR/sub-mm and UV/optical-based measurements. The first set includes the works by \cite{Magnelli2011a,Gruppioni2013a,Magnelli2013a,Casey2012b,Swinbank2014a,Bourne2017a,Koprowski2017a,Liu2018a, Williams2019a,Magnelli2019a,Wang2019b,Lim2020a,Dudzeviciute2020a}, and the second set the results reported in \cite{Wyder2005a,Schiminovich2005a,Robotham2011a,Cucciati2012a,Dahlen2007a,Reddy2009a,Bouwens2012a,Schenker2013a,Finkelstein2015a}. We remind the reader that our model is not a fit to those data points but are shown as a mean for comparison. 

The dust-obscured component traced by the FIR-to-mm surveys has dominated the cosmic history of star formation for the past $\sim12\,$billion years, with a peak era between $z=2-2.5$ ($\sim10-11\,$Gyr ago) and contributing around $\sim80\%$ of total SFRD (see middle panel in Figure \ref{fig:sfrd}). Beyond this peak redshift, the dust-obscured star formation rapidly decreases, following the strong  evolution of the number density of the IR luminosity function (see \S\ref{secc:lum_func}), with values that are comparable to the unobscured star formation traced by the rest-frame UV/optical surveys at $z\sim4$.


At higher redshifts, the dust re-processed star formation rate density  becomes less dominant than the unobscured star formation. This is because of the combination of the flat faint-end slope of the IR luminosity function and the steep redshift evolution of its number density as compared to the UV luminosity function. At $z=5$, the dust-obscured star formation represents $35\%^{+10\%}_{-25\%}$ of the total SFRD and decreases to $25\%^{+15\%}_{-20\%}$ at $z=6$.
Given that massive, IR bright galaxies dominate the obscured component (as shown in the top panel of Figure \ref{fig:sfrd}), the observed decline of the dust-obscured star formation likely reflects the  dearth of massive galaxies at high redshifts.


\subsubsection{Comparison to other measurements and model predictions}

\begin{figure}[t]\hspace{-0.6cm}
\includegraphics[width=0.52\textwidth]{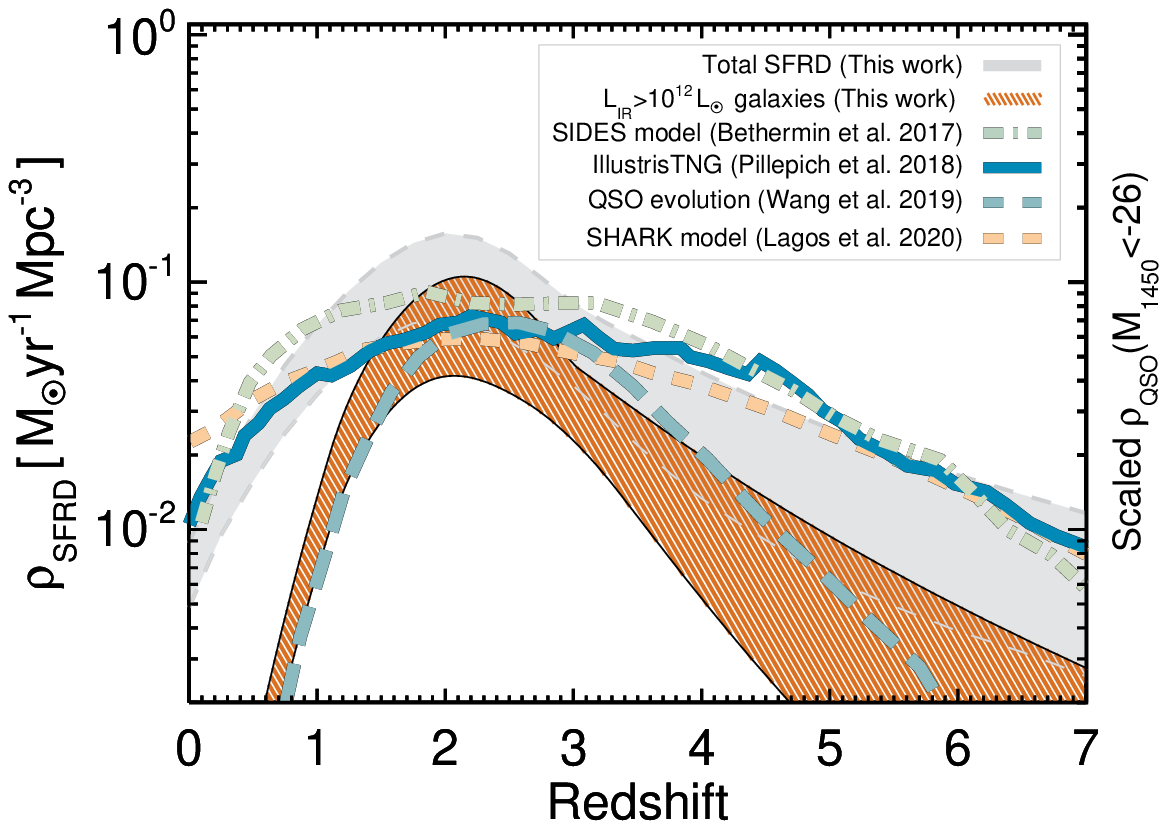}
\caption{The total (obscured $+$ unobscured) SFRD derived in this work (gray region) in comparison to the predictions from galaxy evolution models. The solid blue line represents the predictions from the cosmological hydrodynamical IllustrisTNG simulations (\citealt{Pillepich2018a}), 
the dashed champagne line those from the  {\sc shark} semi-analytic model (\citealt{Lagos2018a}), and the dash-dotted green line the results from the SIDES simulations (\citealt{Bethermin2017a}). Additionally, we include the density evolution of luminous quasars from \citealt{Wang2019c} (scaled for better visualization), which strongly resemble the shape of the SFRD from bright DSFGs with $L_{\rm IR}>10^{12}\,L_\odot$ illustrated by the orange dashed region.
\label{fig:sfrd_models}}
\end{figure}

The SFRD described above is in line with previous results from (sub-)mm surveys, although most of them were limited to $z\lesssim4-5$. For example, \citealt{Dunlop2017a} reported a transition from unobscured-dominated star formation to obscured-dominated at $z\approx4$, in very good agreement with our results. Nevertheless, the small area of their survey ($\sim4.5\,$sq.\,arcmin) prevented them from deriving conclusions beyond this redshift. The results from the larger ALMA surveys presented by \citet{Hatsukade2018a} and \citet{Franco2018a}, covering $26\rm\,arcmin^2$ and $69\rm\,arcmin^2$, respectively, also indicate a minor contribution from DSFGs in the $z\approx4-5$ range (see also \citealt{Yamaguchi2019a}). 
More recently, \citet{Dudzeviciute2020a} used ALMA observations to investigate the properties of $\sim700$ DSFGs detected over $\sim1\,\rm deg^2$, and inferred a SFRD in the range of $\rho_{obs}\approx3\times10^{-3}$ at $z\approx4-5$ for galaxies with $S_{850\mu m}>1\,\rm mJy$ (see also \citealt{Koprowski2017a}), in very good agreement with our results (particularly if we look at the contribution from $L_{\rm IR}>10^{12}\,L_\odot$ galaxies; see Figure \ref{fig:sfrd_models}).


The `wedding cake' structure of all these surveys ensure that the contributions from  the more abundant faint galaxies and the rare bright sources are accounted for (with the possible exception of the most extreme galaxies with ${\rm log}(L_{\rm IR}/L_\odot)\gtrsim13.5$; see Figure \ref{fig:lum_limits}), suggesting that we are not missing any significant population of galaxies. This is also supported by the fact that the number counts predicted by the model are in good agreement with the deepest ALMA observations achieved to-date and with the large-area single-dish telescope surveys (see Figures \ref{fig:num_counts} and \ref{fig:num_counts_short_waves}). 


Other studies based on stacking analysis on (sub-)mm maps and using samples of UV/optically-selected galaxies have also concluded that the fraction of star formation that is obscured by dust decreases at high redshifts (e.g. \citealt{Capak2015a,Bouwens2016a,Fudamoto2020a}). Indeed, \citet{Bouwens2020a} complemented previous results from the literature with their dust-corrected SFRs to estimate the obscured and unobscured components of the cosmic history of star formation, and found that the CSFRD transitions from being primarily unobscured to obscured at $z\sim5$, in relatively good agreement with our estimations\footnote{Note, however, that in \citet{Bouwens2020a} 
most of the dust-obscured star formation at $z>4$ is produced by faint (UV-selected) galaxies with $L_{\rm IR}<10^{12}\,L_\odot$. Nevertheless, those estimations were done assuming SED templates whose dust temperatures increase with redshift. If there is no significant evolution on the dust temperature (e.g. \citealt{Dudzeviciute2020a}) then the SFRs derived for these galaxies would be lower by a factor of $\sim2.5$, as discussed by the authors. This would decrease the contribution of faint galaxies and bring our results into better agreement.} from direct mm-selected samples.

Our results are also compatible with estimations from other tracers such as radio observations. For example, using the VLA-COSMOS 3\,GHz radio survey, \cite{Novak2017a} inferred lower limits for the SFRD up to $z\sim5$ by integrating the radio luminosity function after converting the radio luminosities to SFRs. These VLA-COSMOS 3\,GHz radio constraints  are consistent with our measurements. The authors also provided a completeness-corrected estimation of this quantity by extrapolating the luminosity function to account for the faintest star-forming galaxies. Their estimates show good consistency with our total SFRD, with the possible exception of their last bin at $z\sim5$. Nevertheless, the large extrapolations involved in this process plus the systematic uncertainties that go into calculating SFRD from radio data introduce very large uncertainties in these measurements. The most recent estimates of SFRD from the  VLA-COSMOS 3\,GHz survey presented by \citet{Leslie2020a} are in better agreement with our results.

There are, however, a few other studies that have proposed a different picture, with 
the obscured component significantly dominating the CSFRD back to $z\sim5$. \citet{Rowan-Robinson2016a} estimated the obscured star formation rate density using a sample of {\it Herschel}-selected galaxies over $\sim20\,\rm deg^2$, finding values  far higher than the UV estimations. 
Their constraints, however, cannot rule out the possibility that the UV-based measurements are dominant given the large uncertainties on their estimations, as mentioned by the authors. Additionally, it is possible that their calculations might be contaminated by the effect of gravitationally lensing or by overestimated {\it Herschel} flux densities (in view of their extreme SFRs which extend to $20,000\,M_\odot\,\rm yr^{-1}$).  

More recently, \citet{Gruppioni2020a} 
derive the dust-obscured SFRD using the serendipitously detected sources in the ALPINE survey, finding values in excess to those derived from UV/optical surveys even at $z\sim5$. While the use of highly-confused {\it Herschel} observations might also overestimate the derived luminosities and SFRs, we think that the discrepancy is mainly due because of the possible clustering of serendipitous detections around the original targets. The clustering is expected since the observations' original  targets are massive galaxies (log($M/M_\star$)$\gtrsim10.5$) at $z\approx4-6$. In this case, their measurements would be more representative of an over-dense region in the large-scale structure of the Universe. We note, however, that a significant fraction of their estimates comes from the extrapolation of the IR luminosity function since the 
reported total SFRD is a factor of $\sim5$ greater than the SFRD estimated when using only the detected sources. Therefore, it is also possible that the assumptions on the extrapolation of the IRLF are responsible for part of the observed discrepancy.

In Figure \ref{fig:sfrd_models} we compare the SFRD derived in this work with  results from galaxy evolution models, including the predictions from the cosmological hydrodynamical IllustrisTNG simulations (\citealt{Pillepich2018a}), the {\sc shark} semi-analytic model (\citealt{Lagos2018a}), and the results from the SIDES simulations (\citealt{Bethermin2017a}). Generally speaking, and taking into account the associated uncertainties in these values, our results are in good agreement with the aforementioned studies, pointing towards a convergence between galaxy evolution models and observations.

Finally, we include the density evolution of luminous ($M_{1450}<-26$) quasars (\citealt{Wang2019c}) in  Figure \ref{fig:sfrd_models} (scaled for visualization). The shape of the space density of bright quasar strongly resemble that from the bright DSFGs with $L_{\rm IR}>10^{12}\,L_\odot$ (shaded orange region in the figure), suggesting a connection between these two populations and, therefore, between the onset of star formation and the growth of their massive black holes (e.g. \citealt{Wall2005a}).

\section{Discussion and Conclusions}\label{secc:conclusions}

Exploiting the far-infrared and sub-millimeter data aggregated over the last two decades, and particularly, the state-of-the-art ALMA blind surveys at 1.2\,mm, 2\,mm, and 3\,mm, we have constrained the evolution of the IR luminosity function and dust-obscured star formation in the last 13 billion years, back to $z\sim7$. 
This is achieved  by combining a model of the dusty star-forming galaxy
population with those long-wavelength observations, inferring constraints on the prevalence and
characteristics of these galaxies through measurements of galaxy number counts at different
wavelengths. 


By using a library of SEDs an assuming different evolutionary scenarios for the IR luminosity function, the
model makes predictions for galaxy (sub-)millimeter surveys and then  works backward to discriminate between the different luminosity function scenarios through the  galaxy number counts.  Finally,  the  dust-obscured star formation rate density is estimated by integrating the IR luminosity function (see also \citealt{Casey2018b,Casey2018a,Zavala2018c}).

The model's constraints used in this work include our 2\,mm galaxy number counts 
derived as part of the 2
2\,mm MORA survey, the largest ALMA survey to-date sensitive to detect DSFGs and dust-obscured star formation up to the epoch of reionization and the only one carried out at a 2\,mm wavelength (see also Casey et al. in preparation). Additionally, we use the  3\,mm number counts reported in \S\ref{secc:3mm_num_counts}, which are based on our ALMA  follow-up observations on the sample of 3\,mm-selected galaxies reported by \citet{Zavala2018c}. Finally, we also include the number counts from the ASPECS survey (\citealt{Gonzalez-Lopez2019a,Gonzalez-Lopez2020a}), a deep ALMA large program at 1.2\,mm and 3\,mm.  Altogether, these surveys add a total of  $\sim400\,$sq.\,arcmin. of deep observations with arcsecond-resolution, representing the state-of-the-art blind ALMA surveys to-date.

All these data provide  good constraints on the model parameters, and thus, on the 
IR luminosity function and its evolution with redshift. Based on our best-fit model, which simultaneously reproduces the far-infrared and sub-millimeter data from single-dish telescopes and interferometric surveys, we constrained the faint-end slope of the infrared luminosity function to be flat, with a value of $\alpha_{\rm LF}=-0.42^{+0.02}_{-0.04}$ (see also \citealt{Koprowski2017a,Lim2020a}). This implies that  deep  pencil-beam  observations at (sub-)mm wavelengths would not significantly increase the number of detected sources, in line with previous results from the literature (e.g. \citealt{Popping2020a}). 
The characteristic number density of the luminosity function, $\Phi_\star$, decreases as $\Phi_\star\propto(1+z)^{-6.5^{+0.8}_{-1.8}}$ at $z\gtrsim2$, in a similar fashion as the 
quasar luminosity function and the  density evolution  of  UV-bright galaxies (\citealt{Hopkins2007a,Finkelstein2015a}), which might  suggest  that  these  galaxies  occupy  similar dark matter halos. 

Our constraints on the dust-obscured star formation indicate that the cosmic history of star formation had a peak at $z\approx2-2.5$, and has been dominated by the dust-obscured component during the last 12 billion years, back to $z\sim4$, when the unobscured and obscured contributions were comparable. Beyond this epoch, the dust re-processed star formation rate density was less dominant than the visible star formation, contributing around $35\%^{+10\%}_{-25\%}$ at $z=5$ and $25\%^{+15\%}_{-20\%}$ at $z=6$. This suggests that the bulk ($\gtrsim80\%$)  of  the  star  formation   activity in  the  first  billion years of the Universe was not dust enshrouded. 
Given the massive nature of DSFGs, this drop-off of the obscured component is in line with the decreasing number of high-mass galaxies with increasing redshift (see also \citealt{Dunlop2017a,Bouwens2020a}).\\

Our picture of the history of the cosmic star formation is consistent with previous results from long-wavelength (FIR-to-mm) surveys (e.g. \citealt{Dunlop2017a,Bourne2017a,Hatsukade2018a,Dudzeviciute2020a}), although most of those were limited to $z\lesssim4$. Hence, our results represent a significant progress on our understanding of the prevalence of DSFGs during the first 1.5 billion years of the Universe, and complement the significant efforts carried out using UV/optically selected galaxies (e.g. \citealt{Bouwens2020a}).  
The inferred SFRD is also in broad agreement with the most recent predictions from galaxy evolution models (like IllustrisTNG and {\sc shark}; \citealt{Pillepich2018a,Lagos2020}), 
which point towards a convergence between models and observations.

With estimations for both the obscured and unobscured components, the shape of the cosmic history of star formation is now constrained out to the end of the epoch of reionization. This measurement, which preserves the galaxy mass assembly history 
of the Universe, provides a benchmark against which to compare galaxy formation models and simulations,  and a step forward in our understanding of  the dust and metal enrichment of the early Universe.

\acknowledgments

{\small 
We thank the anonymous referee for a careful review of our manuscript and his/her valuable suggestions.

ALMA is a partnership of ESO (representing its member states), NSF (USA) and NINS (Japan), together with NRC (Canada), MOST and
ASIAA (Taiwan), and KASI (Republic of Korea), in
cooperation with the Republic of Chile. The Joint
ALMA Observatory is operated by ESO, AUI/NRAO
and NAOJ. The National Radio Astronomy Observatory
is a facility of the National Science Foundation operated under cooperative agreement by Associated Universities, Inc. The Dunlap Institute is funded through an endowment established by the David Dunlap family and the University of Toronto.  JAZ and CMC acknowledge thank the University of Texas at Austin College of Natural Sciences for support. CMC also thanks the National Science Foundation for support through grants AST-1714528 and AST-1814034 and support from the Research Corporation for Science Advancement from a 2019 Cottrell Scholar Award sponsored by IF/THEN, an initiative of Lyda Hill Philanthropies. M.A. has been supported by the grant “CONICYT + PCI + INSTITUTO MAX PLANCK DE ASTRONOMIA MPG190030” and “CONICYT+PCI+REDES 190194”. KIC acknowledges funding from the European Research Council through the award of the Consolidator Grant ID 681627-BUILDUP. JH acknowledges support of the VIDI research programme with project number 639.042.611, which is (partly) financed by the Netherlands Organisation for Scientific Research (NWO). KK acknowledges support from the Knut and Alice Wallenberg Foundation. GEM acknowledges  the Villum Fonden research grant 13160 “Gas to stars, stars to dust: tracing star formation across cosmic time” and the Cosmic Dawn Center of Excellence funded by the Danish National Research Foundation under then grant No. 140. E.T. acknowledges support from FONDECYT Regular 1190818, CONICYT PIA ACT172033 and Basal-CATA AFB170002 grants. 

Finally, our deepest gratitude to all the people who continue to provide essential services during the COVID-19 pandemic.\\

{\bf Data and materials availability:}
This paper makes use of the following ALMA data: ADS/JAO. ALMA\#2018.1.00231.S, ADS/JAO.ALMA \#2018.1.00478.S, and ADS/JAO.ALMA\#22019.1.00838.S, archived at https://almascience.nrao.edu/alma-data/archive.
}
%





\bibliography{sample63}
%

\appendix

\begin{figure*}[h!]
\includegraphics[width=\textwidth]{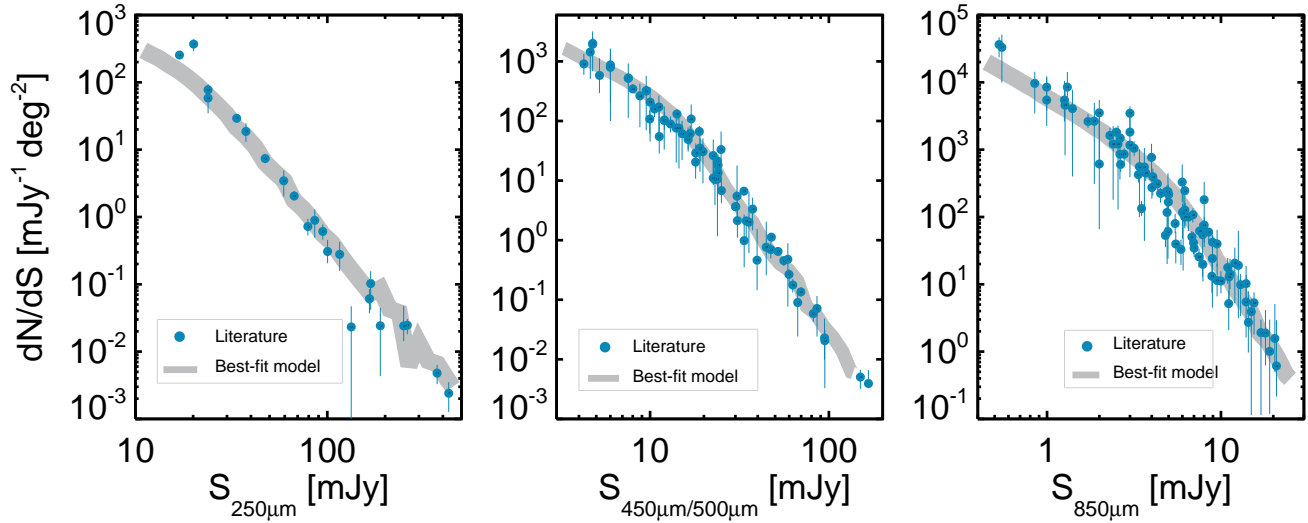}\vspace{0.7cm}
\caption{ Differential number counts at $250\,\mu\rm m$ (left panel), $450\,\mu\rm m/500\,\mu\rm m$ (middle panel), and $850\,\mu\rm m$ (right panel). The predictions from the best-fit model are represented by gray lines, while data points from the literature are illustrated as blue circles. Note that in order to minimize the computational cost, only predictions from a  model with $\alpha_{\rm LF}=-0.42$, $\Psi_2=-6.5$, and $\beta_{\rm em}=1.8$ are included, i.e. without taking into account the uncertainties in the best-fit model parameters. 
Generally speaking, the model reproduces all the far-infrared and sub-millimeter data aggregated over the last two decades from both single-dish telescopes and interferometric surveys spanning more than a decade in wavelength and order of magnitudes in flux density.
At both $250\,\mu\rm m$ and $500\,\mu\rm m$, the data come from BLAST (\citealt{Patanchon2009a,Bethermin2010a}) and {\it Herschel} (\citealt{Oliver2010a,Clements2010a,Bethermin2012a}). The $450\,\mu\rm m$ data points (which have not been scaled relative to the $500\,\mu\rm m$ flux density) come from \citealt{Casey2013a,Chen2013a,Geach2013a, Hsu2016a,Wang2017a,Zavala2017a}. Finally, the $850\,\mu\rm m$ panel includes the works of \citealt{Chapman2002a,Webb2003a,Coppin2006a,Scott2006a,Beelen2008a,Knudsen2008a,Weiss2009a, Casey2013a,Chen2013a,Karim2013a,Hsu2016a,Zavala2017a}.
\label{fig:num_counts_short_waves}}
\end{figure*}

\begin{minipage}{0.95\textwidth}
\section{Number counts model predictions at shorter wavelengths}\label{appendix}

\hspace{0.3cm} As discussed in \cite{Casey2018b,Casey2018a}, the number counts at $\lambda\lesssim850\,\rm\mu m$ only inform about the evolution of the IRLF at $z\lesssim3.0$, where the model parameters are already relatively well understood. Given that our primary goal is to constrain the IRLF at earlier epochs,
only the  number counts at 1.2, 2 and 3\,mm
were used in our analysis. Additionally, 
incorporating all the available number counts in the fitting procedure, particularly those from single-dish telescope surveys, would increase the computational cost significantly since the simulated area would increase by two order of magnitudes. This would make our fitting analysis prohibitive.

\hspace{0.3cm} Nevertheless, as mentioned in the main text, our best-fit model nicely reproduces the number counts at shorter wavelengths, from $\lambda=70\,\mu\rm m$ to $850\,\mu\rm m$, covering a large dynamic range of flux density. In Figure \ref{fig:num_counts_short_waves}, three different examples are shown; the number counts at $250\,\mu\rm m$, $450\,\mu\rm m/500\,\mu\rm m$, and $850\,\mu\rm m$.
The remaining number counts can be found in \cite{Casey2018a}. Since  most of the galaxies detected at these short wavelengths are relatively bright galaxies at $z\lesssim3$, the best-fit parameters studied in this work (which govern the faint and high-redshift population) do not significantly change the model predictions at those wavelengths compared to \cite{Casey2018a}.

\end{minipage}

\newpage
\section{Luminosity limits of the adopted surveys}\label{appendix2}

\begin{figure*}[h!]
\centering
\includegraphics[width=0.63\textwidth]{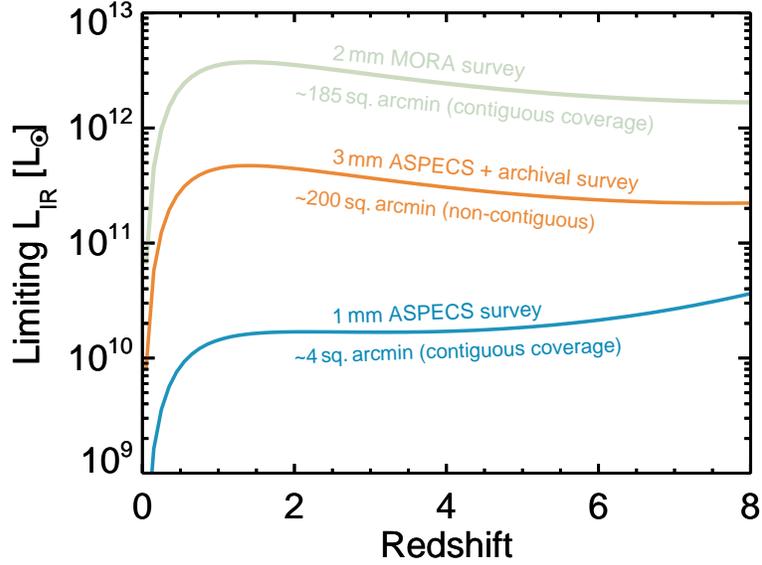}\vspace{0.7cm}
\caption{Luminosity sensitivity limits of the three different surveys adopted in this work. 
Given that we use directly the galaxy number counts to constrain our model, these luminosity limits were calculated adopting the minimum flux density bin of each of the galaxy number counts (see Figure \ref{fig:num_counts}), respectively. The deepest number counts at 1.1\,mm are sensitive to galaxies down to $L_{\rm IR} \sim2\times10^{10}\,l_\odot$ up to $z\sim8$, while the larger area surveys at 2\,mm and 3\,mm are sensitive to brighter and rarer galaxies with $L_{\rm IR}>10^{11}-10^{12}$. Note that, although  the cosmic SFRD presented in Figure \ref{fig:sfrd} was calculated by integrating the best-fit IR luminosity function over the interval log$(L/L_\odot):[9,13.8]$ (i.e. extrapolating beyond the surveys' luminosity limits), the majority of the contribution arise from galaxies with  luminosities in the range of $L_{\rm IR}=10^{11}-10^{13}\,L_\odot$, well within the luminosity limits of the adopted data. 
\label{fig:lum_limits}}
\end{figure*}



\end{document}